\documentclass[aps,prd,amsmath,preprintnumbers,nofootinbib,a4paper,preprint,floatfix]{revtex4-1}

\usepackage{longtable}
\usepackage{amsthm}
\usepackage{amssymb}
\usepackage{amsfonts}
\usepackage{bbm}
\usepackage{color}
\usepackage{dcolumn}
\usepackage{bm}
\usepackage{pxfonts}
\usepackage{slashed}
\usepackage{graphicx}
\usepackage{multirow}

\usepackage[
  bookmarks]{hyperref}
  
\newcommand{\beq}{\begin{eqnarray}}
\newcommand{\eeq}{\end{eqnarray}}

\newcommand{\bmp}{\noindent\begin{minipage}{16cm}}
\newcommand{\emp}{\end{minipage}\vskip 7mm} 

\def\wick#1{\setbox2=\hbox{$\displaystyle#1$}
    \setbox3=\null\ht3=3.0pt\dp3=0.0pt\wd3=20.0pt
    #1\kern-\wd2\kern3.0pt\raise11.0pt\vbox{\hrule height0.3pt
    \hbox{\vrule width0.3pt\box3\vrule width0.3pt}}\kern-24.0pt\kern\wd2}

\def\longwick#1{\setbox2=\hbox{$\displaystyle#1$}
    \setbox3=\null\ht3=3.0pt\dp3=0.0pt\wd3=27.0pt
    #1\kern-\wd2\kern3.0pt\raise11.0pt\vbox{\hrule height0.3pt
    \hbox{\vrule width0.3pt\box3\vrule width0.3pt}}\kern-31.0pt\kern\wd2}

\def\verylongwick#1{\setbox2=\hbox{$\displaystyle#1$}
    \setbox3=\null\ht3=3.0pt\dp3=0.0pt\wd3=43.0pt
    #1\kern-\wd2\kern3.0pt\raise11.0pt\vbox{\hrule height0.3pt
    \hbox{\vrule width0.3pt\box3\vrule width0.3pt}}\kern-47.0pt\kern\wd2}

\definecolor{bluc}{cmyk}{1,1,0,0.1}
\definecolor{rossoCP3}{cmyk}{0,.88,.77,.40}
\definecolor{rosso}{cmyk}{0,1,1,0.4}
\definecolor{rossos}{cmyk}{0,1,1,0.55}
\definecolor{rossoc}{cmyk}{0,1,1,0.2}
\definecolor{verdes}{cmyk}{0.92,0,0.59,0.4}

\hypersetup{colorlinks, bookmarksopen, bookmarksnumbered,
citecolor=verdes, linkcolor=bluc, pdfstartview=FitH, urlcolor=rossos}

\theoremstyle{definition}

\theoremstyle{plain}

\def\lsim{\mathrel{\rlap{\lower4pt\hbox{\hskip1pt$\sim$}}
    \raise1pt\hbox{$<$}}}                
\def\gsim{\mathrel{\rlap{\lower4pt\hbox{\hskip1pt$\sim$}}
    \raise1pt\hbox{$>$}}}                

\newcommand{\TeV}{\,{\rm TeV}}
\newcommand{\GeV}{\,{\rm GeV}}

\newcommand{\ba}{\begin{eqnarray}}
\newcommand{\ea}{\end{eqnarray}}
\newcommand{\be}{\begin{equation}}
\newcommand{\ee}{\end{equation}}
\newcommand{\bd}{\begin{displaymath}}
\newcommand{\ed}{\end{displaymath}}
\newcommand{\een}{\nonumber\end{equation}}
\newcommand{\bea}{\begin{eqnarray}}
\newcommand{\eean}{\nonumber\end{eqnarray}}
\newcommand{\eea}{\end{eqnarray}}

\def\eq#1{Eq.~(\ref{#1})}

\def\fig#1{Fig.~\ref{#1}}

\def\tab#1{Table~\ref{#1}}

\def\cite#1{\citep{#1}}

\newcommand{\fm}{\mathrm{fm}}

\newcommand{\mps}{m_{\rm{PS}}}
\newcommand{\fps}{F_{\rm{PS}}}

\newcommand{\mv}{m_{\rm{V}}}
\newcommand{\ma}{m_{\rm{A}}}

\newcommand{\mpcac}{m_{\rm{PCAC}}}
\newcommand{\mf}{m_{\rm{f}}}

\def\mcA{{\mathcal A}}
\def\mcB{{\mathcal B}}
\def\mcC{{\mathcal C}}

\def\mcE{{\mathcal E}}

\def\mcO{{\mathcal O}}

\def\la{\langle}
\def\ra{\rangle}
\def\psibar{\overline{\psi}}

\def\tr#1{{\mathrm{tr}\left[#1\right]}}

\newcommand{\old}[1]{}

\begin{document}

\title{\Large  \color{rossoCP3} SU(2) Gauge Theory with Two Fundamental Flavours:\\ a Minimal Template for Model Building }
\author{Rudy Arthur$^{\color{rossoCP3}{\varheartsuit}}$}\email{arthur@cp3-origins.net}
\author{Vincent Drach$^{\color{rossoCP3}{\spadesuit}}$}\email{vincent.drach@cern.ch}
\author{Martin Hansen$^{\color{rossoCP3}{\varheartsuit}}$}\email{hansen@cp3-origins.net}
\author{Ari~Hietanen$^{\color{rossoCP3}{\varheartsuit}}$}\email{hietanen@cp3-origins.net}
\author{Claudio Pica$^{\color{rossoCP3}{\varheartsuit}}$}\email{pica@cp3-origins.net}
\author{Francesco Sannino$^{\color{rossoCP3}{\varheartsuit}}$}\email{sannino@cp3-origins.net}

\affiliation{
\vspace{5mm}
{$^{\color{rossoCP3}{\varheartsuit}}${ \color{rossoCP3}  \rm  CP}$^{\color{rossoCP3} \bf 3}${\color{rossoCP3}\rm-Origins} \& the {\color{rossoCP3} \rm Danish IAS}, University of Southern Denmark, Campusvej 55, DK-5230 Odense M, Denmark \\}
{$^{\color{rossoCP3}{\spadesuit}}$ Theoretical Physics Department, CERN, Geneva, Switzerland}
\vspace{5mm}
 }
\begin{abstract}
We investigate the continuum spectrum of the SU(2) gauge theory with $N_f=2$ flavours of fermions in the fundamental representation.
This model provides a minimal template which is ideal for a wide class of Standard Model extensions featuring novel strong dynamics that range from composite (Goldstone) Higgs theories to several intriguing types of dark matter candidates, such as the SIMPs. 
We improve our previous lattice analysis~\cite{Hietanen:2014xca} by adding more data at light quark masses, at two additional lattice spacings, by determining the lattice cutoff via a Wilson flow measure of the $w_0$ parameter, and by measuring the relevant renormalisation constants non-perturbatively in the RI'-MOM scheme.
Our results for the lightest isovector states in the vector and axial channels, in units of the pseudoscalar decay constant, are  $m_V/\fps\sim 13.1(2.2)$ and $m_A/\fps\sim 14.5(3.6)$ (combining statistical and systematic errors). 

In the context of the composite (Goldstone) Higgs models, our result for the spin-one resonances are  $m_V > 3.2(5)~\TeV$ and $m_A > 3.6(9)~\TeV$, which are above the current LHC constraints.
In the context of dark matter models, for the SIMP case our results indicate the occurrence of a compressed spectrum at the required large dark pion mass, which implies the need to include the effects of spin-one resonances in phenomenological estimates. 
 \end{abstract} 
 \preprint{CP3-Origins-2016-006 DNRF90}
 \preprint{DIAS-2016-6}
 \preprint{CERN-TH-2016-037}
\maketitle

\section{Introduction}

New composite dynamics is often invoked to construct extensions of the Standard Model (SM) physics that can address one or several of the SM shortcomings. 

For example, composite extensions have been suggested to replace the SM Higgs sector, to suggest natural dark matter (DM) candidates and, more recently, to explain \cite{Molinaro:2015cwg,Matsuzaki:2015che,Franzosi:2016wtl} the  observed tantalizing excess in the diphoton decay  channel or earlier diboson excesses \cite{Fukano:2015hga,Franzosi:2015zra} recorded by the CMS and ATLAS experiments~\cite{CMS:2015dxe,ATLAS-2015,Aad:2015owa}.     Time-honoured classes of fundamental electroweak composite dynamics are Technicolor (TC) \cite{Weinberg:1975gm,Susskind:1978ms} and composite Goldstone Higgs models \cite{Kaplan:1983fs,Kaplan:1983sm}.
 
In TC models the Higgs boson is the lightest scalar excitation of the fermion condensate responsible for electroweak (EW) symmetry breaking~\cite{Sannino:1999qe,Hong:2004td,Dietrich:2005jn,Dietrich:2005wk,Sannino:2009za}. 
The physical Technicolor Higgs mass can be light due to near conformal dynamics \cite{Sannino:1999qe,Dietrich:2006cm} and the interplay between the TC sector and the SM fermions and electroweak gauge bosons~\cite{Foadi:2012bb}. 

In composite Goldstone Higgs models~\cite{Kaplan:1983fs,Kaplan:1983sm}, the new sector has an underlying fundamental dynamics with larger global symmetry group than the one strictly needed to break  the EW symmetry. 
In this case the Higgs state can be identified with one of the additional Goldstone Bosons (GB), and it is therefore naturally light. 
However, to break the EW symmetry, typically radiative corrections are not enough and yet another sector is required to induce the correct vacuum alignment for the EW gauge bosons and for the Higgs to acquire the observed mass.

The underlying fundamental theory studied here constitutes the ultra minimal composite template for any natural UV completion that simultaneously embodies both the TC and composite Goldstone Higgs models~\cite{Appelquist:1999dq,Ryttov:2008xe,Galloway:2010bp,Cacciapaglia:2014uja}. It is also well known that fermion mass generation constitutes a challenge for any composite dynamic extension. For the present theory an extension that makes use of chiral gauge theories \cite {Appelquist:2000qg,Shi:2014yxa,Shi:2015fna} has been put forward recently in \cite{Cacciapaglia:2015yra}. The constructions yield distinctive experimental signatures and can be used universally for both types of model building.

Novel composite dynamics has also been advocated to construct natural candidates for DM stemming from a composite EW sector.
In fact, several asymmetric DM candidates were put forward which are stable baryons in TC models~\cite{Nussinov:1985xr,Barr:1990ca} or Goldstone bosons of a new strong sector~\cite{Gudnason:2006ug,Gudnason:2006yj,Ryttov:2008xe,Frandsen:2011kt}.  

Another interesting class of DM models, unrelated to the composite EW scenario,  was recently revived in  \cite{Hochberg:2014dra}. Here an alternative mechanism \cite{Carlson:1992fn,deLaix:1995vi} is employed for achieving the observed DM relic density.  
It uses $3\to2$ number-changing processes that should occur in the dark sector involving strongly interacting massive particles (SIMPs).  
Compared to the WIMP paradigm, where the dark matter particles typically are expected to be around the TeV scale, this model can yield dark matter particles with masses of a few 100 MeVs. In \cite{Hochberg:2014kqa,Hansen:2015yaa} a realisation of the SIMP paradigm was introduced in terms of composite theories for which the model investigated here again provides the minimal template.  
In this realisation, the pions constitute the dark matter particles and the topological  Wess-Zumino-Witten (WZW) term \cite{Wess:1971yu,Witten:1983tw,Witten:1983tx}  introduces a 5-point pion interaction, making it an ideal candidate for the $3\to2$ annihilation process.  
The most minimal realisation of this breaking pattern comes indeed from the underlying Sp(2)=SU(2) gauge group (but can be generalised to any Sp(N$_c$) gauge group). 
The first consistent investigation of the phenomenological viability of this construction, that properly takes into account important next-to-next-leading-order corrections via chiral perturbation theory, appeared in  \cite{Hansen:2015yaa}.  
Here it was  shown that higher order corrections substantially increase the tension with phenomenological constraints. 
Because the energy scale of the SIMP is very light, it is especially relevant to know at which energy scale dark spin-one resonances will appear, or more generally to understand its spectroscopy \cite{Hochberg:2015vrg}. 
Furthermore the new states will modify the scattering at higher energies introducing possible interesting resonant behaviours \cite{Choi:2016hid} and, as it is the case for ordinary QCD, will impact on a number of dark-sector induced physical observables.

 
In this work we investigate the SU(2) gauge theory with $N_f=2$ flavours of Dirac fermions in the fundamental representation. 
One important feature of this minimal SU(2) template model is that, due to the pseudo-reality of the fundamental representation, the flavour symmetry is upgraded to an SU(4) (locally isomorphic to SO(6)) symmetry which is expected to break spontaneously to Sp(4) (locally isomorphic to SO(5)), thus leading to 5 Goldstone bosons. 

The theory has previously been studied on the lattice, and in particular, it has been shown that the expected pattern of spontaneous chiral symmetry breaking is realised~\cite{Lewis:2011zb}. 
A first estimate, affected by large systematic errors, of the masses of the vector and axial-vector mesons, in units of the pseudoscalar meson decay constant, have been obtained in~\cite{Hietanen:2014xca}. 
The scattering properties of the Goldstone bosons of the theory have also been considered~\cite{Arthur:2014zda}, and the model has furthermore been investigated in the context of possible DM candidates related to the EW scale~\cite{Hietanen:2013fya,Drach:2015epq}.
Other groups have also investigated the spectrum of this model on the lattice \cite{Hayakawa:2013maa,Amato:2015dqp} concluding that chiral symmetry is broken, although no continuum extrapolation was attempted as the focus of both works was on the comparison with the six flavours theory to understand the approach to the conformal window in $SU(2)$ gauge theories.

Here we extend our previous analyses by improving our control on the systematics. Our simulations achieve smaller fermion masses,  include two additional lattice spacings, and we also perform a precise determination of the lattice spacings used. Finally we determine the relevant renormalisation constants non-perturbatively.
 
The paper is organised as follows.  
We first describe the lattice setup in section \ref{sec:setup} and the procedure to set the lattice spacing through the Wilson Flow observable $w_0$ in section \ref{sec:WF}. In section~\ref{sec:NPR} we discuss the calculation of the renormalisation constants using the RI'-MOM scheme. Finally we provide in section~\ref{sec:spectro} an improved estimation of the spectrum of the theory.

\section{Lattice set-up}\label{sec:setup}

We simulate the SU(2) gauge theory with two Dirac fermions in the fundamental
representation discretised using the (unimproved) Wilson action for two mass-degenerate fermions $u$, $d$ and the Wilson plaquette action for the gauge field.
The numerical simulations have been performed using an improved version of the HiRep code first described in \cite{DelDebbio:2008zf}.
The fermionic part of the action reads:
 \begin{eqnarray}
S_F &=&  \sum_x\overline{\psi}(x)(4+am_0)\psi(x) \nonumber \\
   && - \frac{1}{2}\sum_{x,\mu}\left(\overline{\psi}(x)(1-\gamma_\mu)U_\mu(x)\psi(x+\hat\mu)
   +\overline{\psi}(x-\hat\mu)(1+\gamma_\mu)U_\mu^\dagger(x)\psi(x)\right) \,,
\end{eqnarray}
where $U_\mu$ is the gauge field,  $\psi$ is the doublet of $u$ and $d$ fermions, and $am_0$ is
the $2\times2$ diagonal mass matrix proportional to the identity.

Our simulation are performed at four values of the inverse lattice gauge coupling $\beta$, for various fermion masses and on
several volumes. This is needed in order to perform the required extrapolations to the chiral limit and infinite volume and to give an estimate of the systematic errors stemming from such extrapolations. We detail the procedure used in the following sections.

The bare parameters of our simulations are summarised in Table~\ref{table:sim_param}. 
We have extended our previously published dataset considerably, in particular towards the chiral regime and by adding two additional lattice spacings at $\beta=1.8, 2.3$. As we will discuss in more detail below, note that the lightest quark masses now reach, in some cases, the decay threshold for the vector meson resonance.
The simulations in Table~\ref{table:sim_param} denoted with an asterisk, are only used to study the systematic errors due to finite size effects. The remaining runs will be referred to as ``large volume runs'' in this paper.
This is justified as for all these lattices we have $\mps\, L\ge 5$ which implies a systematic error of  about $5\%$ for the quantities studied here~\cite{Hietanen:2014xca}. 

\begin{table}[t]
  \begin{tabular}{ccc}
    \hline
    $\beta$ & Volume & $a m_0$  \\
    \hline
    \hline
    1.8 & $16^3\times32$~~  & -1.00, -1.089, -1.12, -1.14,
    -0.15, -1.155$^\ast$ \\
     1.8 & $32^3\times32$~~  & -1.155  -1.557\\
    \hline
    2.0 & $16^3\times32$~~  & -0.85, -0.9, -0.94,
    -0.945, -0.947$^\ast$  \\
    2.0 & $32^4$ &  -0.947, -0.949, -0.-952,-0.957,-0.958 \\
    \hline
    2.2 & $16^3\times32$  & -0.60, -0.65, -0.68
    -0.70, -0.72$^\ast$, -0.75$^\ast$ \\
    2.2 & $24^3\times 32$ &  -0.75$^\ast$  \\
    2.2 & $32^4$ & -0.72,-0.735, -0.75  \\
    2.2 & $48^4$ & -0.76  \\
    \hline
    2.3 & $32^4$ & -0.575,-0.60,-0.625,-0.65,-0.675, -0.685 \\
    \hline
  \end{tabular}
  \caption{Parameters used in the simulations. Runs with $^\ast$ are used only to study finite size effects. All the others runs are referred to in the text as ``large volume runs''.\label{table:sim_param}}
\end{table}

For convenience, we define the following operators:
\begin{eqnarray}
{\cal O}_{\overline{u}d}^{(\Gamma)}(x) &=& \overline{u}(x)\Gamma d(x) \,, 
\end{eqnarray}
where $\Gamma$ denotes any product of Dirac matrices. 

We extract the meson masses from zero-momentum two-point correlation functions
\begin{align}
f_{\Gamma}(t)
 & =  \sum_{\vec x} \left\langle {\cal O}_{\overline{u}d}^{(\Gamma)}(t,\vec x\,)^\dagger {\cal O}_{\overline{u}d}^{(\Gamma)}(0) \right\rangle.
\end{align}

The quantities of interest in this study are the
pseudoscalar $\Gamma=\gamma_5$, vector $\Gamma=\gamma_k$ ($k=1,2,3$),
and axial vector $\Gamma=\gamma_0\gamma_5\gamma_k$ mesons.  We use  $Z_2\times Z_2$ single time slice stochastic sources~\cite{Boyle:2008rh} to estimate the meson 2-point correlators. From those, we define an effective mass $m_{(\Gamma)}^{\rm{eff}}(t)$ as  in~\cite{DelDebbio:2007pz,Bursa:2011ru} by the solution of the implicit equation:
\be\label{eq:meff}
\frac{f_{\Gamma}(t-1)}{f_{\Gamma}(t)} = \frac{e^{-m_{(\Gamma)}^{\rm{eff}}(t) (T-(t-1))} + e^{-m_{(\Gamma)}^{\rm{eff}}(t) (t-1)}}{e^{ -m_{(\Gamma)}^{\rm{eff}}(t)(T-t)} + e^{-m_{(\Gamma)}^{\rm{eff}}(t) t }}\,,
\ee
where $T$ is lattice time extent. At large Euclidean time, $m_{(\Gamma)}^{\rm{eff}}(t)$ approaches the value of the mass of the lightest state with the same quantum numbers as the operator ${\cal O}_{\overline{u}d}^{(\Gamma)}$.
In the following, we will denote the pseudoscalar meson mass $\mps$, and the isovector vector and axial-vector  meson mass $\mv$ and $\ma$ respectively.

In addition to the meson masses above, we will use in the present analysis two other
quantities: the current quark mass $\mpcac$ and the Goldstone boson decay
constant $\fps$. We define the quark mass through the Partially Conserved Axial Current (PCAC)
relation
\begin{equation}                                                                                     
  \mpcac=\lim_{t \rightarrow \infty}\frac{1}{2}\frac{\partial_t f_{AP}(t)}{f_{\gamma_5}(t)},         
\label{eq:PCAC}                                                                                      
\end{equation}                                                                                       
where                                                                                                
\begin{align}                                                                                        
  f_{AP} (t) &= \sum_{\vec x} \left\langle {\cal O}_{\overline{u}d}^{(\gamma_0\gamma_5)}(t,\vec x\,)^\dagger {\cal O}_{\overline{u}d}^{(\gamma_5)}(0) \right\rangle \, .
\end{align}                                                                                                                                                       
The Goldstone boson decay constant can be calculated as:
\begin{equation}                                        
 \fps = \frac{2 \mpcac}{\mps^2} G_{\rm PS},
\end{equation}
where $G_{\rm PS}$ is obtained from the asymptotic form of $f_{\gamma_5}(t)$ at large $t$:
\begin{equation}
  f_{\gamma_5}(t) \sim -\frac{G_{\rm PS}^2}{\mps}\exp\left[-\mps t\right].
\end{equation}
On a lattice of finite temporal extent, we use the same definitions as in~\cite{DelDebbio:2007pz,Bursa:2011ru}.

The (bare) values in lattice units for $\mpcac$, $\mps$, $\fps$, $\mv$ and $\ma$ corresponding to the large volume lattices considered in this paper are reported in table~\ref{table:results} in appendix \ref{ap:res}. 

To convert the lattice quantities to physical units, we determine the lattice spacing for our
simulations and the appropriate non-perturbative renormalisation constants. 

It is well known that for Wilson fermions, the pseudoscalar decay constant renormalises multiplicatively with the scale independent renormalisation constant $Z_A$  and that the bare PCAC mass renormalises with the ratio $Z_A/Z_P(\mu^2)$.

The lattice spacing, in a generic composite model, is fixed by the requirement that the
renormalised Goldstone boson decay constant has a given value specified for the physical model considered. For example in the case of composite dynamics at the electroweak scale, a value of 246 GeV yields the correct mass for the electroweak gauge bosons. 
For the more general fundamental composite Goldstone Higgs scenario described in \cite{Cacciapaglia:2014uja} the scale is still set by the same requirement, but the constraint on the renormalised Goldstone boson decay constant now reads $\fps\, \sin (\theta)=246$ GeV. 
The actual value of the parameter $\theta$ in this model depends on the electroweak gauge bosons corrections, the top corrections as well as the effects of other possible sources of explicit breaking of the initial SU($4$) symmetry. 
The Technicolor limit is recovered for $\theta = \pi/2$ while the composite pGB Higgs case corresponds to small, but non-vanishing $\theta$. 
Any other value of $\theta$ is also allowed and the resulting model thus interpolates between these two limits. 
For the details we refer to~\cite{Cacciapaglia:2014uja}. 

Another case of immediate interest is the SIMPlest composite model \cite{Hochberg:2014kqa} for DM where, as shown in \cite{Hansen:2015yaa}, it is important to control the underlying dynamics. By stretching chiral perturbation theory to its limit of validity, the interesting phenomenological values for the pion decay constant would be as low as 10 MeV with pion masses of the order of 100 MeV. Besides the rescaling the pion decay constant, another major difference, when compared to composite dynamics at the electroweak scale, resides in the fact that the SIMP requires quite massive pions.

For definiteness, below we present our results in units of the EW scale with $\sin(\theta) = 1$ but the dependence on $\theta$ can be reinstated when needed. At the end we will also comment on the results for the SIMPlest case.

\section{Scale setting}\label{sec:WF}
Following~\cite{Luscher:2010iy}, we consider the following ``Wilson flow'' equation for the gauge fields:
\begin{equation}
\frac{d}{dt} V_t(x,\mu)= -g_0^2 \,\{\partial_{x,\mu}S_{G}(V_t)\}\, V_t(x,\mu)\quad {\rm with} \quad V_{t=0}(x,\mu)=U(x,\mu)\,,
\label{wflow}
\end{equation}
where $t$ denotes the fictitious flow ``time'', $U(x,\mu)$ are the gauge links, and $S_G$ is the plaquette gauge action. 
One important property is that correlation functions at flow time $t>0$ are finite, when the four-dimensional theory is renormalised as usual, and the flow thus maps gauge fields into smooth, renormalised gauge fields~\cite{Luscher:2011bx}. Observables at non-zero flow time can, in particular, be used to define a scale, as shown in~\cite{Luscher:2010iy}.

Two different scale-setting observables have been introduced in the literature, known as $t_0$~\cite{Luscher:2010iy} and $w_0$~\cite{Borsanyi:2012zs}. In terms of $E(t)$, the action density at flow time $t$, they are defined through the following equations:
\bea
 \mcE(t)= t^2 E(t)\,,&&\,\,\,  \mcE(t_0)=\mcE_{\rm{ref}}\,, \\
W(t)= t\frac{d}{dt} \mcE(t)\,,&& \,\,\, W(w_0^2)= W_{\rm{ref}}\,,
\eea
where $\mcE_{\rm{ref}}$ and $W_{\rm{ref}}$ are two dimensionless
reference values. In this work we will use $w_0$ to set the scale.
The value of $w_0$ obtained for each quark mass needs to be extrapolated to the chiral limit to obtain a scale $w_0^\chi$ for each lattice spacing.

We investigated finite volume errors in $w_0$ at the chosen reference value $W_{\rm ref}$ by comparing two simulations performed on spacial sizes $L=16$ ($\mps\, L \sim 5.1$) and $L=32$  ($\mps\, L \sim 8.4$) at bare parameters $m_0=-0.75$ and $\beta=2.2$ . These values of the bare parameters were chosen to correspond to one of the lightest points in our dataset, at a fine lattice spacing. The values of $w_0$ obtained are $w_0(L=16) = 3.39(6)a$ and $w_0(L=32) = 3.36(10)a$ which agree well within statistical errors, indicating that finite volume effect for $w_0$ can be safely neglected for $\mps\, L > 5$ within our numerical precision.

\subsection{Determination of $w_0^{\chi}$}
\begin{figure}[t]
  \centering
\begin{minipage}{.48\textwidth}
  \centering
 \includegraphics[width=\linewidth]{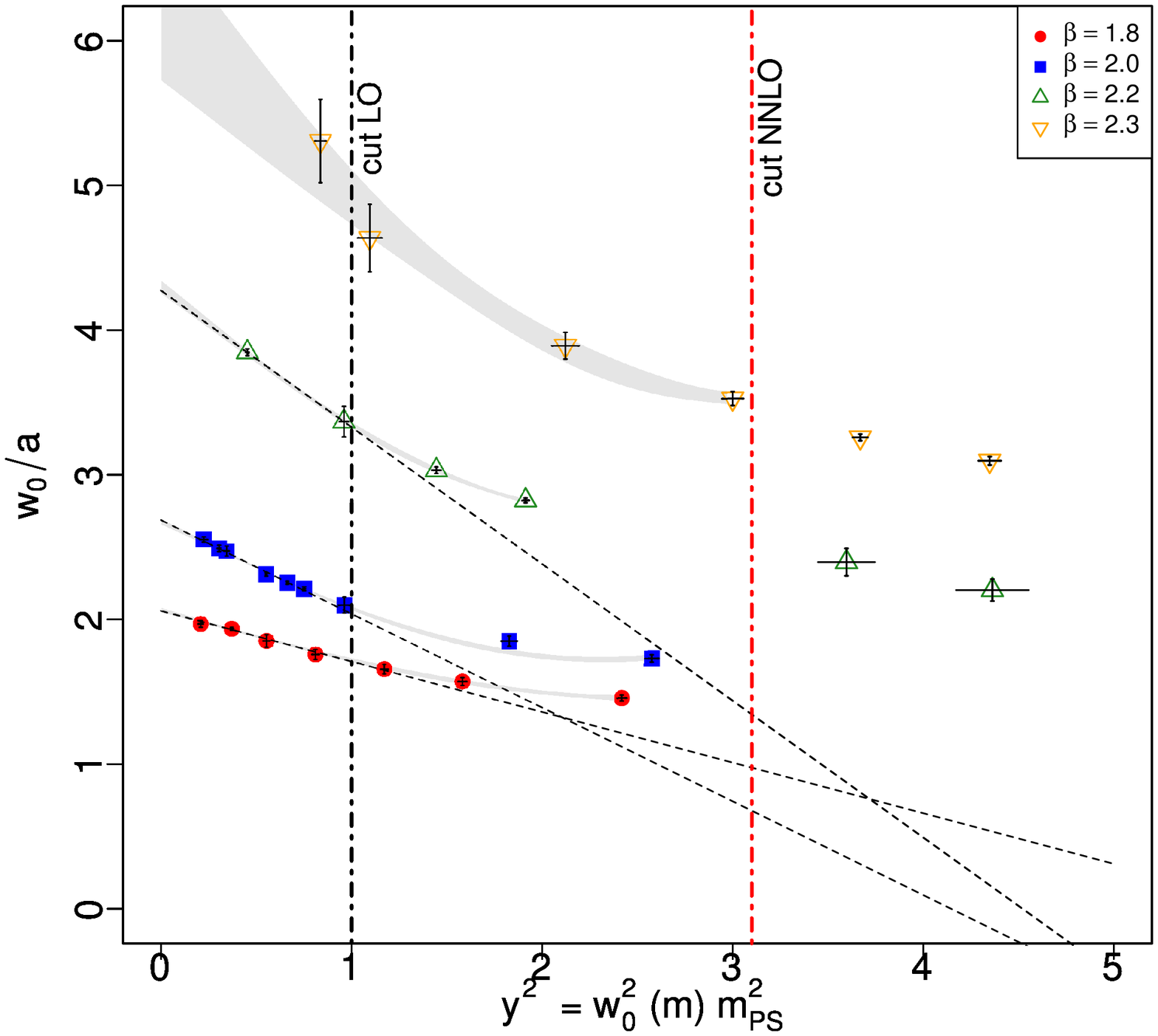}
\end{minipage}%
\hspace*{0.5cm}\begin{minipage}{.48\textwidth}
\centering
  \includegraphics[width=\linewidth]{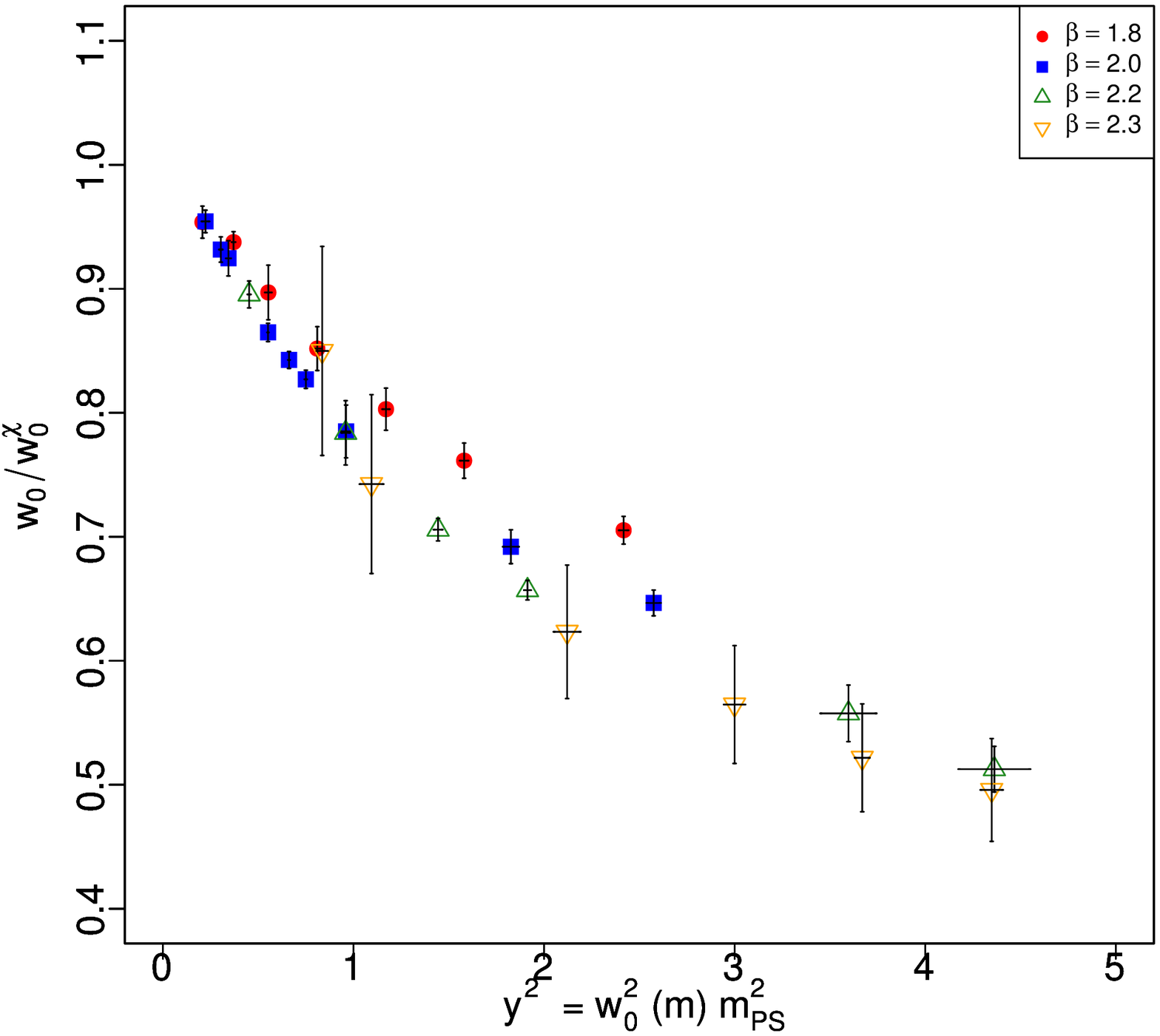}
\end{minipage}
\caption{Chiral behaviour of $w_0$ as a function of $y^2$ in unit of the lattice spacing (left panel) and in unit of $w_0^{\chi}$ (right panel) for $W_{\rm{ref}}=1$. The data at four lattice spacings are displayed. \label{fig:w0_vs_y2} }
\end{figure}

In \fig{fig:w0_vs_y2} (left panel) we show our results for $w_0/a$ for the four lattice spacings considered in this study as a function of $y^2$, where $y=w_0(\mpcac)\,\mps$. 
Here the reference value chosen is $W_{\rm{ref}} =1$.  For all the points in  \fig{fig:w0_vs_y2} we have $\mps\,L > 5.5$ and are thus safe from finite volume effects.

In order to extrapolate to the chiral limit, we use the NNLO expansion in terms of $\mps^2$ which reads~\cite{Bar:2013ora}:
\be
w_0(\mps^2) = w^\chi_0 \left( 1 + k_1 \frac{m_{\rm{PS}}^2}{(4\pi F)^2} +  k_2
\frac{m_{\rm{PS}}^4}{(4\pi F)^4} \log \frac{m_{\rm{PS}}^2}{\mu^2} \right)\,,
\ee
where $F$ is the pseudoscalar decay constant and $k_1$, $k_2$ are dimensionless low energy constants. Note that the chiral logarithm enters only at NNLO. In practice we fitted our data at each $\beta$ with the following ansatz :
\be
w_0(\mps^2)= w^\chi_0 \left( 1 + A y^2 + B y^4\log y^2\right)\,,
\ee
where $A$, $B$ and $w^\chi_0$ are free parameters with the choice $w^\chi_0 \mu = 1$.

\begin{table}[b!]
  \begin{tabular}{ccccc}
    \hline
    $\beta$ & $w_0^{\chi}/a$   & $A$  & $B$  & $\chi^2/\rm{ndof}$   \\
    \hline
    1.8  & 2.066(16) &-0.169(12) &0.022(6) &5.5/4    \\
    2.0  & 2.675(20) &-0.223(10) &0.036(4) &14.8/6  \\
    2.2  & 4.311(49) &-0.224(12) &0.036(8) &1.0/1   \\
    2.3  & 6.202(477) &-0.205(39) &0.018(9) &2/1  \\
    \hline
  \end{tabular}
  \caption{Summary of the NNLO fits for $w_0/a$ for each
    value of the lattice spacing. We chose $W_{\rm{ref}}=1$ as reference value.\label{table:w0_a_NNLO_fit}}
\end{table}

The fit is performed for each of the four $\beta$-values independently and the gray bands indicate the $1\sigma$ error regions. The best fit parameters and their statistical errors are reported in \tab{table:w0_a_NNLO_fit}. In the left panel of \fig{fig:w0_vs_y2}, the red dotted vertical line indicates the upper limit of the $y^2$ region used in the NNLO fit.

For three of our data sets we have also performed a fit to the NLO expression. The black vertical dotted line indicates the upper limit of the $y^2$  region included in the NLO fit. Due to lack of data, we cannot perform this fit for $\beta=2.3$. For the three remaining lattice spacings available, the results of the NLO and NNLO fits agree well within uncertainties.

In the right panel of \fig{fig:w0_vs_y2} we show $w_0/w_0^{\chi}$  for all four lattice spacings. The  deviation from a universal curve of such a quantity is a measure of lattice discretisation errors. As can be seen, these are small in the $w_0$ observable for our three finest lattice spacings. The same conclusion can also be reached by looking at the dimensionless coefficient $A$ and $B$ as determined from the fits, given in  \tab{table:w0_a_NNLO_fit}.


\section{Non-perturbative renormalisation constants}\label{sec:NPR}

\subsection{RI'-MOM scheme}
In this section we describe the method used to determine the non-perturbative renormalisation constants of the isovector vector (V), axial (A), and pseudoscalar (P) bilinear operators. They are needed for the renormalisation of the pseudoscalar decay constant $\fps$ and of the quark mass $\mpcac$.

We use the RI'-MOM scheme (regularisation invariant momentum scheme) as in~\cite{Martinelli:1994ty}.
We define the following bilinear operators :
\be
O_{\Gamma}(x) = \psibar(x) \tau^3 \Gamma \psi(x),\quad\text{with}\quad\Gamma \in \left\{P,V,A,S\right\} \equiv \left\{\gamma_5,\gamma_\mu,\gamma_5\gamma_\mu,1\right\}\, ,
\ee
and the fermion propagator :
\be
S(x,y) =\la \psi(x) \psibar(y) \ra,\quad\text{and}\quad S(p) = \sum_{p} e^{ip(x-y)} S(x,y).
\ee
Note that we have omitted to write explicitly spin and color indices. We also define the following Green's function:
\be
G_\Gamma(p)= \la \psi(p) O_{\Gamma}(p) \psibar(p) \ra
\ee
and we will denote the corresponding vertex function by:
\be\label{eq:vertex_function}
\Pi_\Gamma(p)= S(p)^{-1} G_\Gamma(p) S(p)^{-1},
\ee
where $S^{-1}(p)$ is the inverse propagator in spin and color space.
The RI'-MOM scheme~\cite{Martinelli:1994ty} is then defined by imposing the conditions that in the chiral limit and at a given scale $p^2=\mu^2 $, the inverse propagator and amputated Green's function $\Pi_\Gamma(p)$ satisfy the following equations:
\be
Z_q^{-1} \frac{-i}{4 N_c} \tr{\frac{\gamma_\mu \sin(ap_\mu)}{\sin^2 (ap_\mu)} S^{-1}(p)}\Bigg|_{p^2 = \mu^2} = 1,\quad\text{and}\quad Z_q^{-1} Z_\Gamma  \frac{1}{4 N_c}\tr{P_\Gamma \Pi_\Gamma(p)}\Big|_{p^2 = \mu^2} = 1\,,
\ee
where the trace is over spin and color indices and  the projectors $P_\Gamma$ are defined as follows:
\beq
P_\Gamma \in \left\{ P_P,P_V,P_A,P_S\right\} \equiv \left\{ \gamma_5, \frac{\gamma_{\mu}}{4}, \frac{\gamma_{\mu}\gamma_5}{4},1\right\}\,.
\eeq
For convenience we define:
\beq
\Lambda_q(p^2) &=& \frac{-i}{4 N_c} \tr{\frac{\gamma_\mu \sin(ap_\mu)}{\sin^2 (ap_\mu)} S^{-1}(p)}\,,\\
\Lambda_\Gamma(p^2) &=& -i \frac{\tr{\frac{\gamma_\mu \sin(ap_\mu)}{\sin^2 (ap_\mu)} S^{-1}(p)}}{\tr{P_\Gamma \Pi_\Gamma(p)} }\,, \\
\Lambda_{P/S}(\mu^2)&=& \Lambda_P(p^2)/\Lambda_S(p^2)\,,
\eeq
such that in the chiral limit:
\be
\Lambda_q(\mu^2) = Z_q(a,\mu^2),\quad\Lambda_\Gamma(\mu^2) = Z_\Gamma(a,\mu^2)\quad \text{and}\quad  \Lambda_{P/S}(\mu^2)= Z_P(a,\mu^2)/Z_S(a,\mu^2)\,.
\ee

\subsection{Evaluation of the correlators}
\label{subsec:evaluation_mom_source}
Following the approach introduced in~\cite{Gockeler:1998ye}, we use momentum sources. This approach has the advantage to be computationally inexpensive and to have a high statistical accuracy. We will shortly summarise the procedure.

The vertex functions defined in \eq{eq:vertex_function} are not gauge invariant, and must be computed in a fixed gauge. We chose the Landau gauge by minimising a functional proposed in~\cite{Vink:1992ys}.

We introduce $S(y,p)$ defined to be the solution of the following linear equation
\be
\sum_y D (x,y) S(y,p) =  \mathbbm{1}\, e^{ipx}\,,
\ee
where $\mathbbm{1}$ stands for the identity matrix in spinor and color indices.
It is straightforward to obtain that
\be
G_\Gamma(p) = \frac{1}{V} \sum_z \gamma_5\, S'(z,p)^\dagger   \gamma_5\, S'(z,p),\quad\text{where}\quad S'(z,p)= e^{-ipz}S(z,p)\label{eq:Gp}
\ee
and
\be
S(p) = \frac{1}{V} \sum_x e^{-ipx} S(x,p)\,.\label{eq:Sp}
\ee

\begin{figure}[p!]
  \centering
\begin{minipage}{.48\textwidth}
  \centering
 \includegraphics[width=\linewidth]{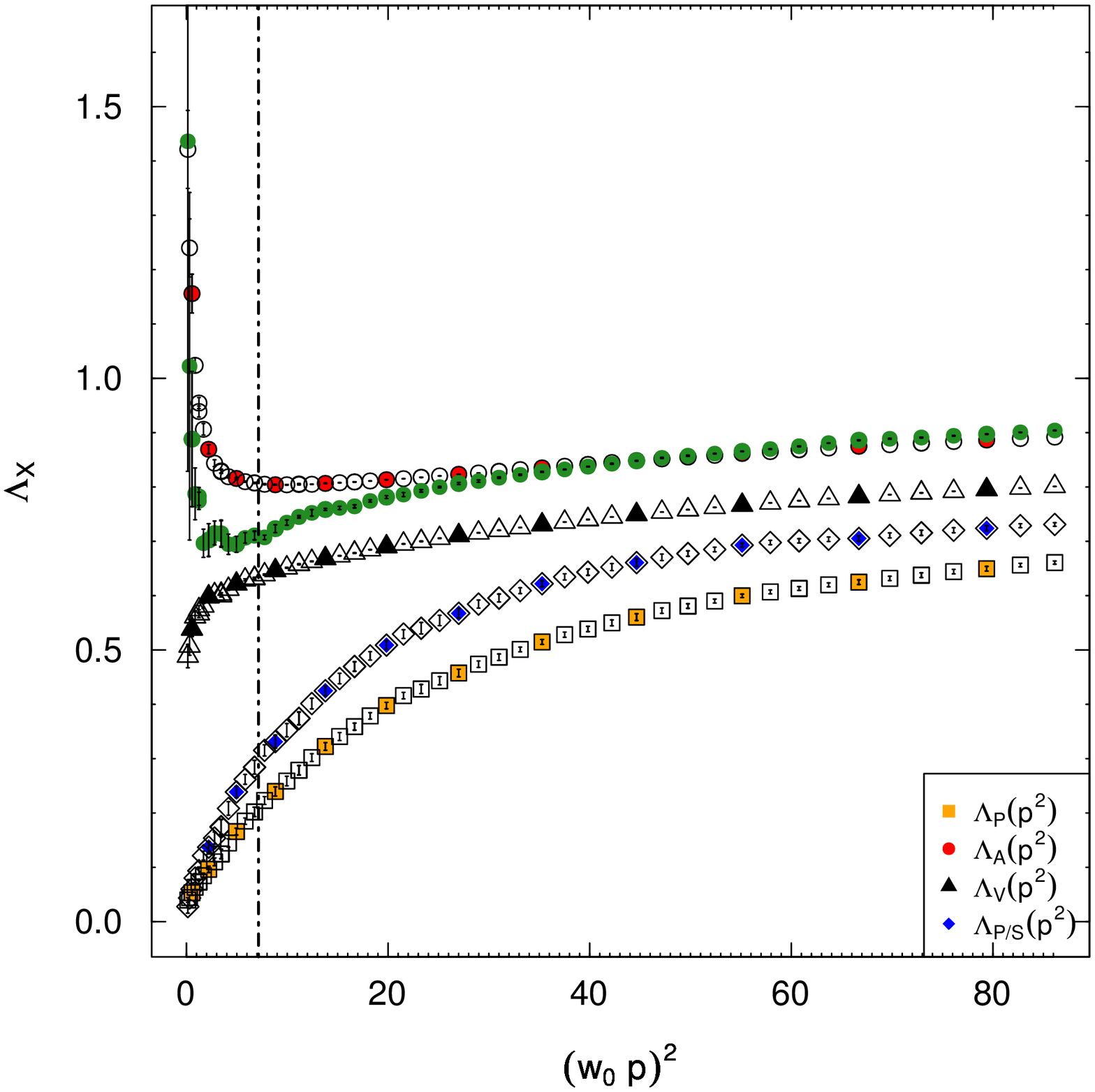}
\end{minipage}%
\hspace*{0.5cm}\begin{minipage}{.48\textwidth}
\centering
  \includegraphics[width=\linewidth]{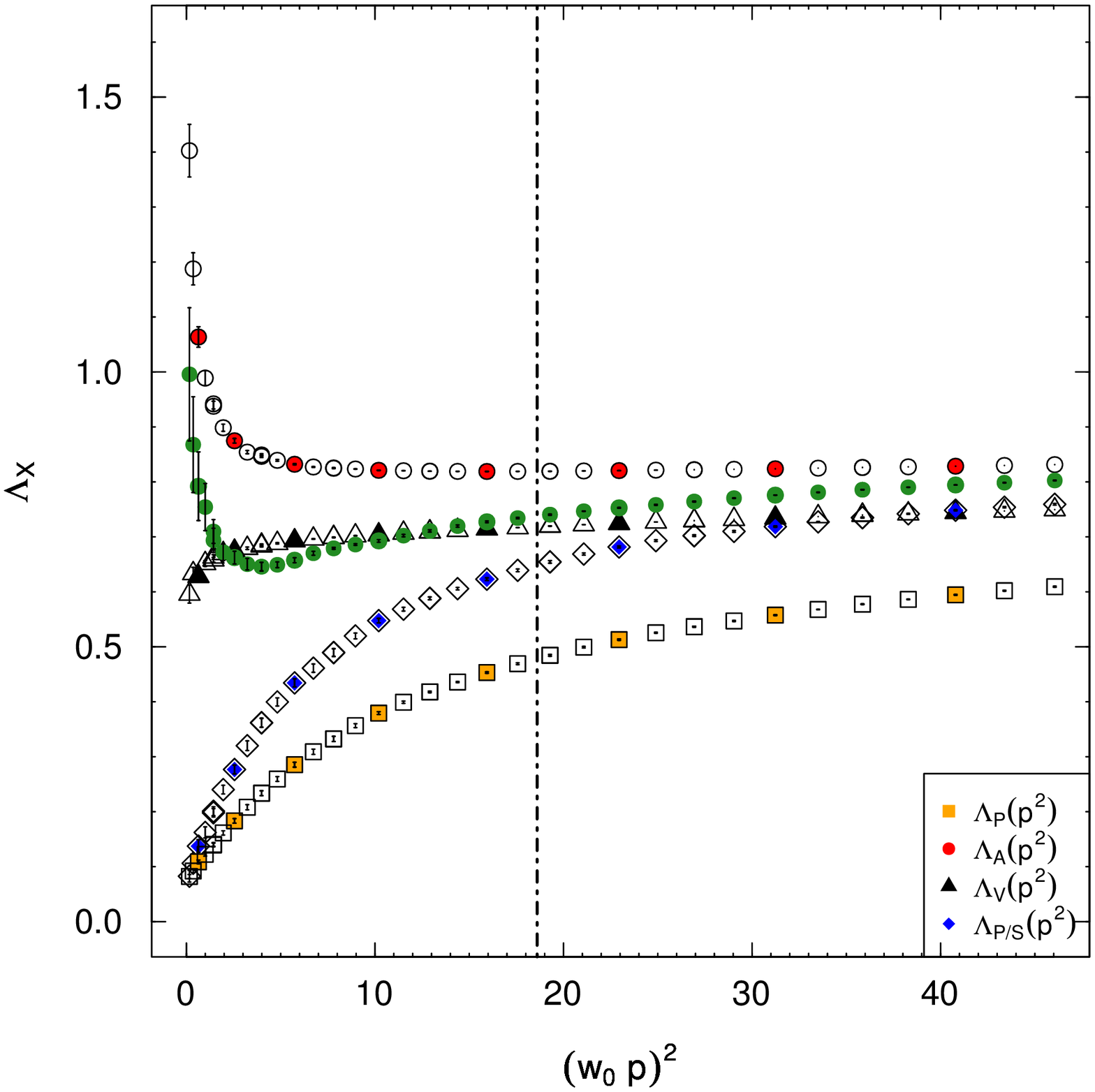}
\end{minipage}
\caption{\label{fig:LambdaX_vs_p2}$\Lambda_{X=P,V,A,P/S}$ as a function of $(ap)^2$  for the most chiral point at $\beta=2.0$ (left panel) and $\beta=2.2$ (right panel). The filled data points are obtained without twisted boundary conditions while the empty symbols denote the use of a non vanishing $\theta$. The vertical lines indicate where $(ap)^2 =1$.}
\end{figure}

\begin{figure}[p!]
  \centering
\begin{minipage}{.48\textwidth}
  \centering
 \includegraphics[width=\linewidth]{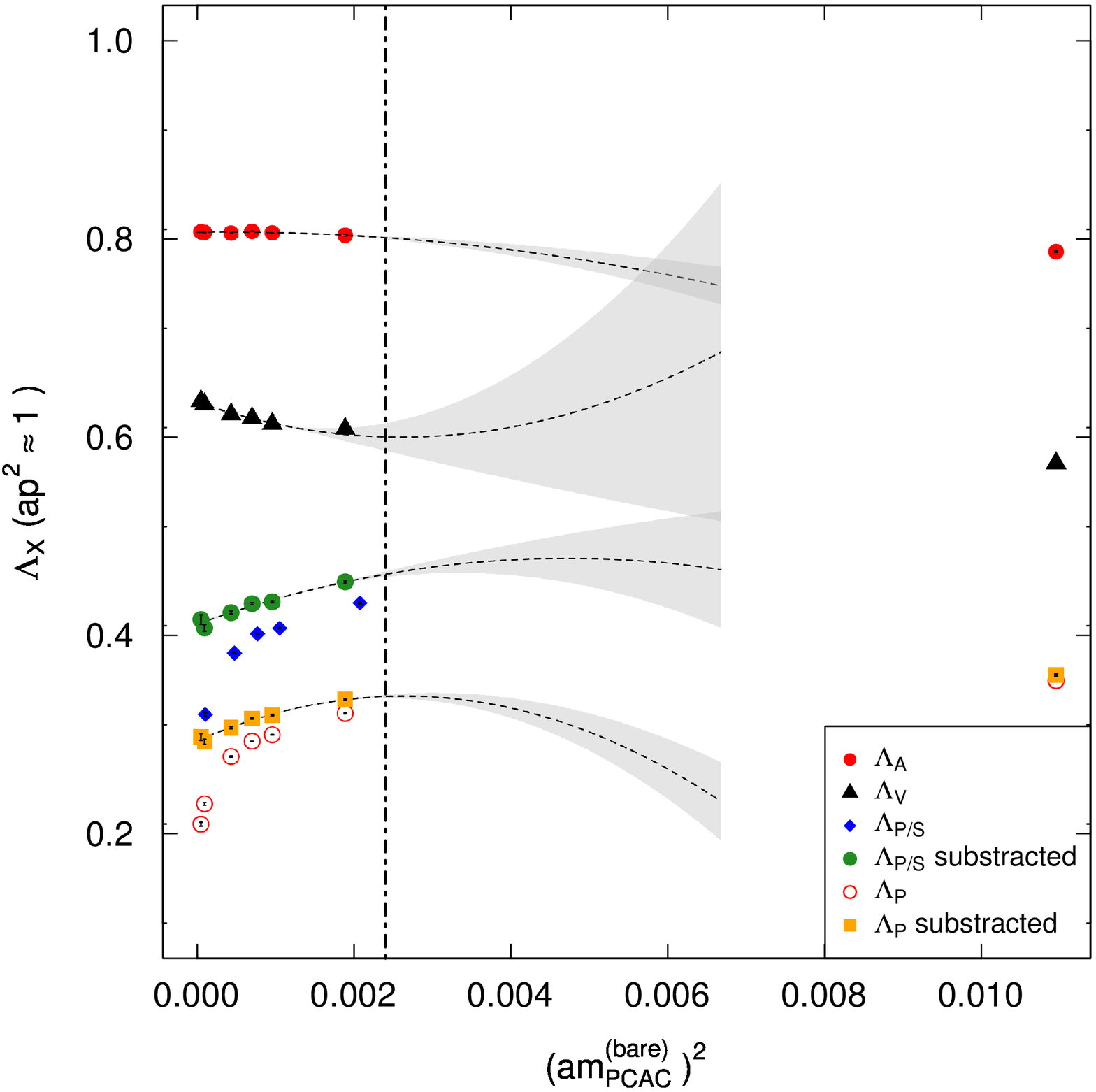}
\end{minipage}%
\hspace*{0.5cm}\begin{minipage}{.48\textwidth}
\centering
  \includegraphics[width=\linewidth]{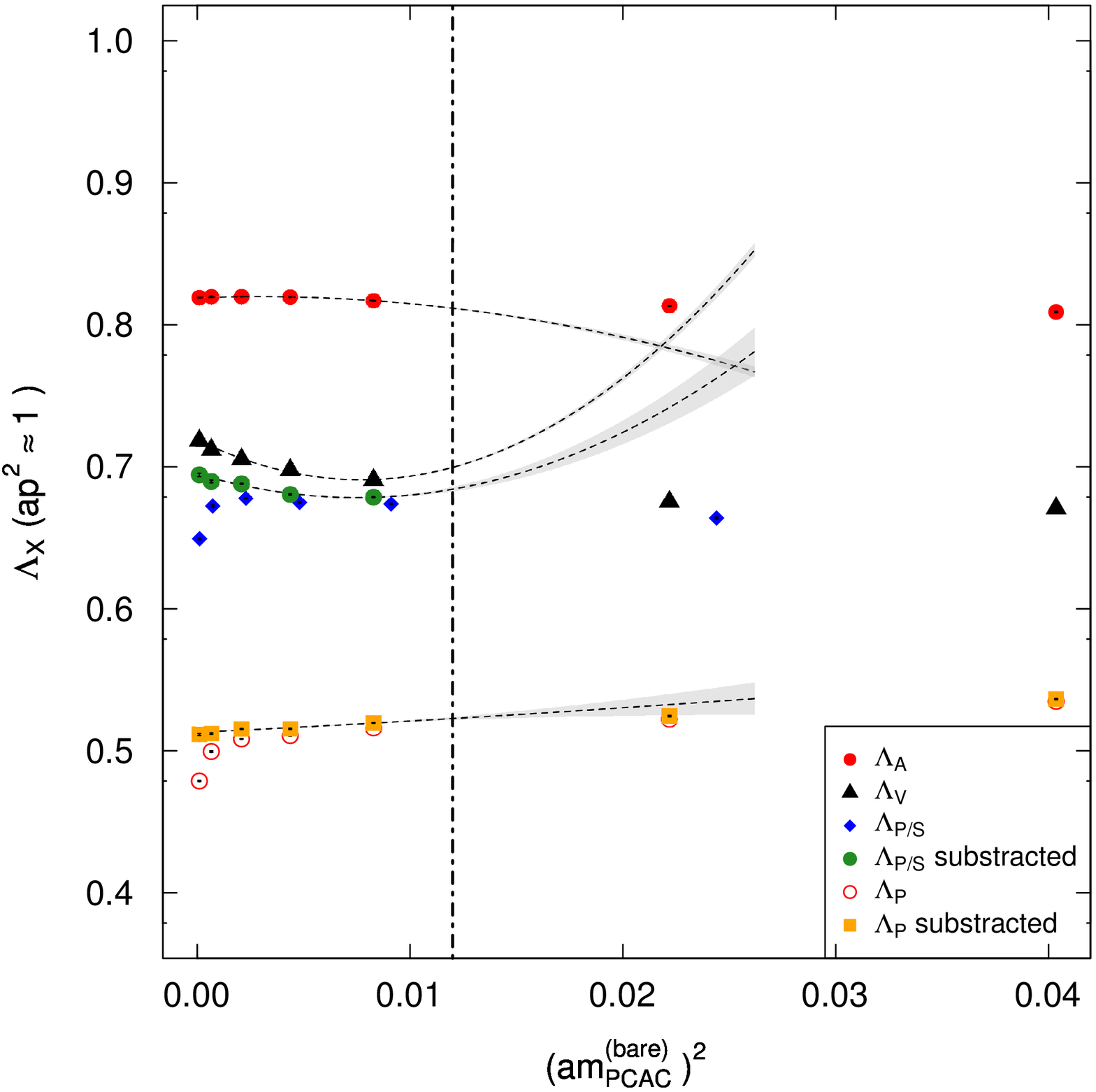}
\end{minipage}
\caption{\label{fig:LambdaX_vs_mpcac}$\Lambda_{X=P,V,A,P/S}$ at fixed $(ap)^2 = 1$ as a function of $(a\mpcac^{\rm{bare}})^2$  for
  $\beta=2.0$ (left panel) and $\beta=2.2$ (right panel). }
\end{figure}

\subsection{Twisted boundary conditions}

In order to interpolate easily between the lattice momenta we use twisted boundary conditions~\cite{Bedaque:2004kc,Sachrajda:2004mi} by imposing :
\be\label{eq:tw_BC}
q(x+L) = e^{iBx} q(x)\quad\text{with}\quad B_\mu = \frac{\pi \theta_\mu}{L_{\mu}}\,,
\ee
where $L_{\mu=1,2,3} = L$ and $L_4=T$ and $\theta$ is the twist angle. The boundary conditions are imposed by modifying the Dirac operator in the valence only. The accessible momenta are then $p_\mu = \frac{2\pi}{L_{\mu}} n_\mu +  \frac{\pi }{L_{\mu}} \theta_\mu$. Eq.~(\ref{eq:Gp}) and (\ref{eq:Sp}) of sect.~\ref{subsec:evaluation_mom_source} can be generalised in the case of twisted boundary conditions.

In practice the propagator $S(p)$  and the Green's function $G_\Gamma(p)$ are evaluated for
\beq
n_{\mu} = l \left(1,1,0,0\right)\quad\text{and}\quad \theta_\mu = l' \frac{1}{2}\left(1,1,0,0\right)\,,\label{eq:twist}
\eeq
for every pair $(l,l')$ with $l \in [\![ 1,\dots,l_{\rm{max}}]\!]$ and $l' \in [\![ -l'_{\rm{max}},\dots,l'_{\rm{max}}]\!]$. Note that we also use negative values for $l'$ in order to obtain the same values of $p^2$ from twisting with different initial momentum. This is useful in order to estimate cut-off effects. From \fig{fig:LambdaX_vs_p2} it is clear that they are small. Finally, note that we choose ``non-democratic'' momenta in \eq{eq:twist}.

\subsection{Results \& Analysis}

The vertex functions $\Lambda_X$ for $X\in\{P,V,A,P/S\}$ at a fixed quark mass, as a function of momentum $(ap)^2$, are shown for  $\beta =2.0$ and $\beta=2.2$ in \fig{fig:LambdaX_vs_p2}. The filled symbols are obtained with twist angle $\theta=0$, while the empty symbol denotes the results obtained for $\theta\neq 0$.

In order to determine the value of the renormalisation constants, the first step is to extrapolate the result in the chiral limit.
At fixed $p^2$ the behaviour of the vertex functions, which do not involve the pseudoscalar density, is expected to be polynomial in $(a\mpcac^{\rm{bare}})^2$. Concerning the pseudoscalar vertex functions, it is well known that special care must taken due to the presence of the Goldstone bosons pole~\cite{Martinelli:1994ty,Cudell:1998ic,Cudell:2001ny}. In that case, we use the following ansatz to perform pion-pole subtraction :
\beq
\Lambda_P(p^2) = \mcA(p^2) + \mcB(p^2) \mpcac + \frac{\mcC(p^2)}{\mpcac}\,,
\eeq
where $\mcA,\mcB$ and $\mcC$ are functions of $p^2$. The subtraction is performed for each $p^2$ by fitting the data at a given $\beta$, and we will denote $\Lambda^{\rm{sub}}_P(p^2) = \Lambda_P(p^2) -\frac{\mcC(p^2)}{\mpcac}$ the subtracted vertex function at a given fermion mass.

\begin{figure}[t!]
  \centering
\begin{minipage}{.48\textwidth}
  \centering
 \includegraphics[width=\linewidth]{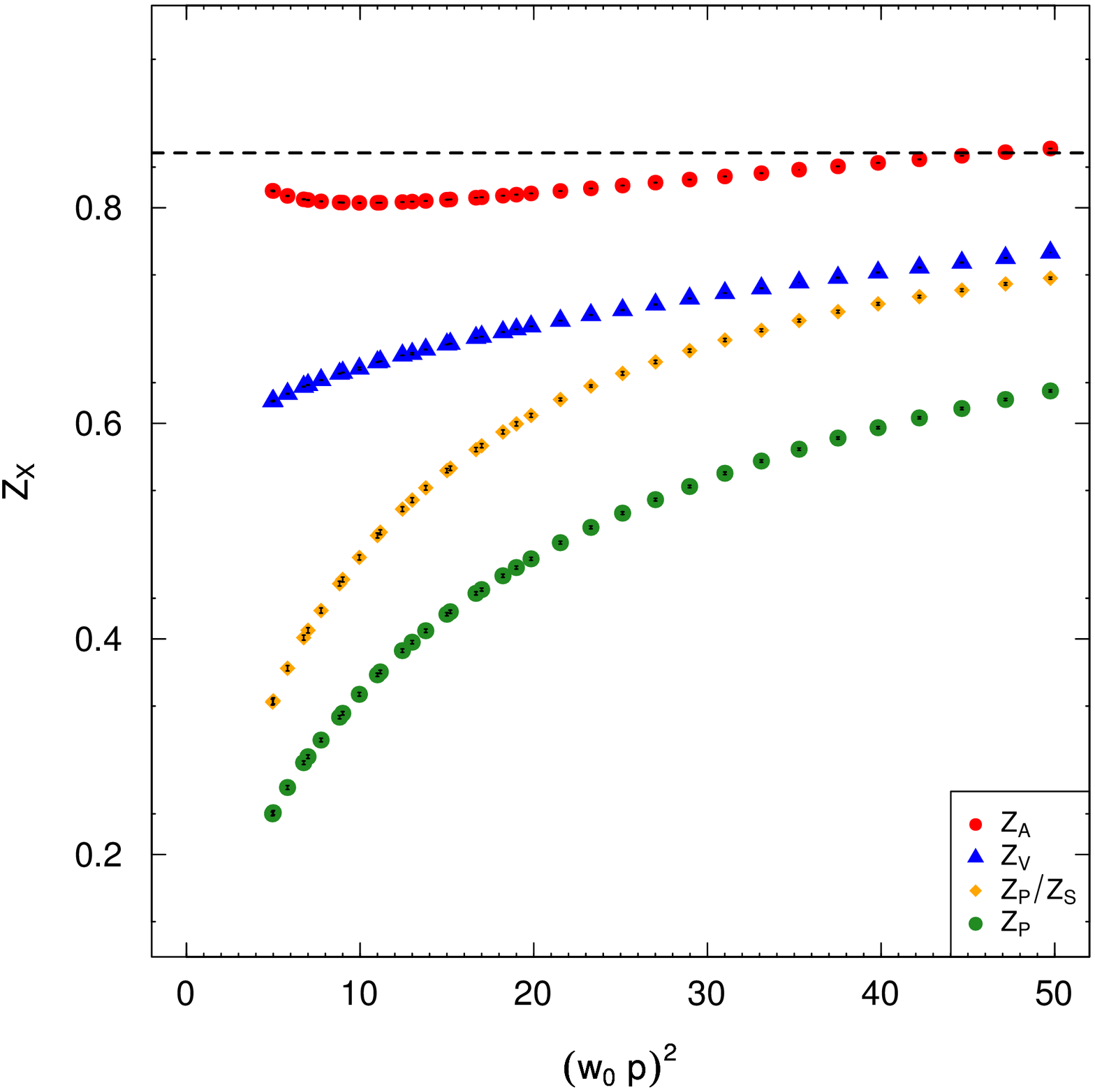}
\end{minipage}%
\hspace*{0.5cm}\begin{minipage}{.48\textwidth}
\centering
  \includegraphics[width=\linewidth]{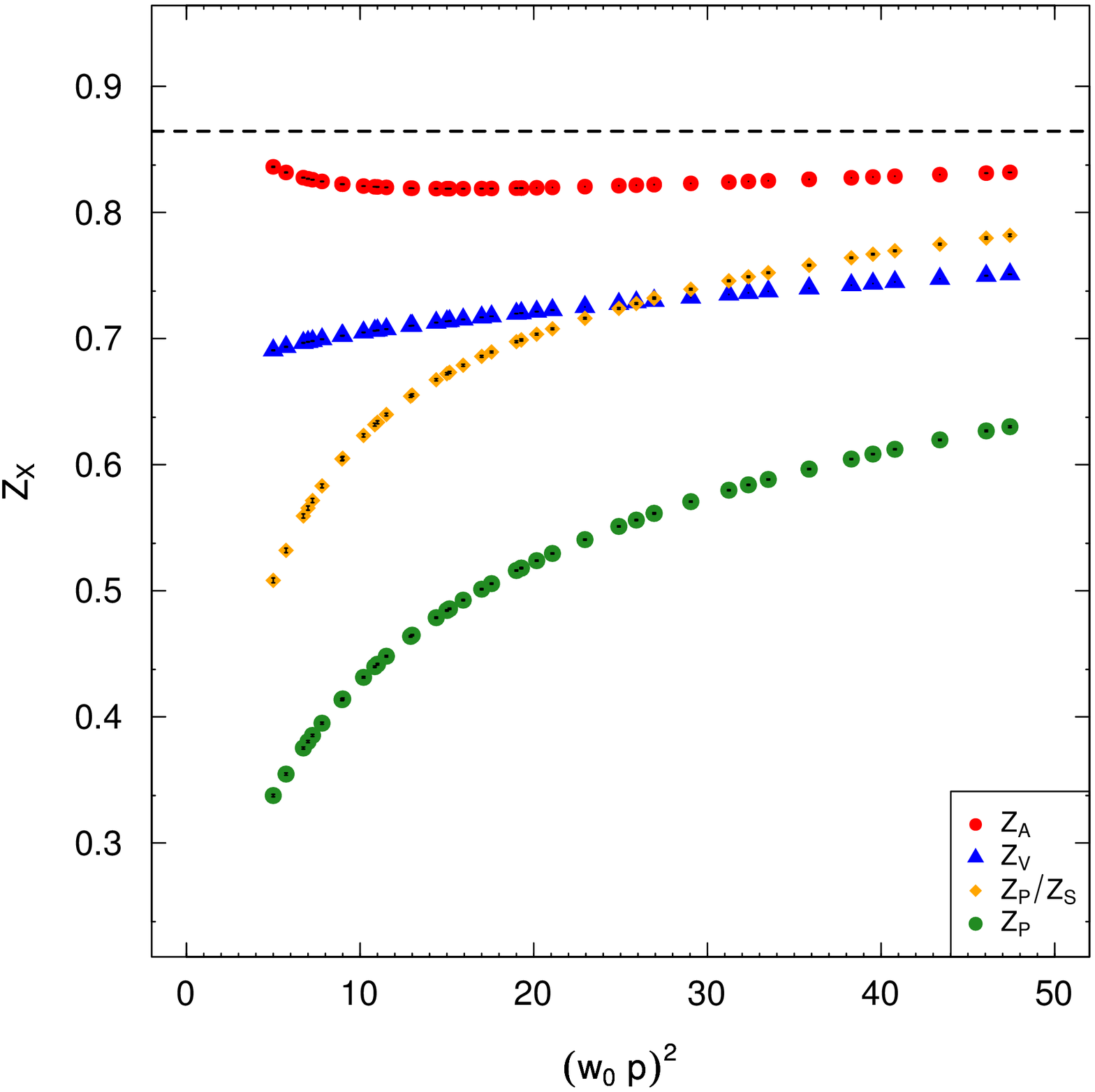}
\end{minipage}
\caption{\label{fig:Zs_vs_p2}$Z_{X=P,V,A,P/S}$  as function of the  renormalisation scale  $ (w_0\mu)^2 = (w_0 p)^2$ for
  $\beta=2.0$ (left panel) and $\beta=2.2$ (right panel).}
\end{figure}

We illustrate the chiral extrapolation at fixed $p^2$ in \fig{fig:LambdaX_vs_mpcac}, where we show $\Lambda_X(p^2=1/a^2)$ as a function of $(a\mpcac^{\rm{bare}})^2$. 
In the plot we also included the Goldstone boson subtracted vertex function for $X=P$ and $P/S$. The chiral extrapolation is obtained by fitting a second order polynomial in $(a\mpcac^{\rm{bare}})^2$ to the data. 
The vertical dashed-dotted line indicates the extent of the region included in the fit. 
The best fit curve and its statistical error are included in the figure. The typical $\chi^2/\rm{ndof}$ for these fits are larger than one, because of the small statistical error bars on $\Lambda_X$. 
Given our target accuracy of a few percent, those effects are negligible, however.

We show in \fig{fig:Zs_vs_p2} the dependence of the chirally extrapolated vertex function $\Lambda_X$ as a function of $(w_0 p)^2$. In the continuum, $Z_V$, $Z_A$ and $Z_P/Z_S$ are renormalisation scale independent, the observed scale dependence is a manifestation of discretisation effects.\footnote{Note that we do not subtract perturbative $\mcO(a)$ effects, and we do not convert $Z_P(\mu^2)$ to the $\overline{MS}$ scheme.}

In order to have meaningful estimates of $Z_X(p^2)$, one relies on the existence of a renormalisation window: $\Lambda < p < \mcO(a^{-1})$. The lower bound guarantees that the Goldstone pole contamination is small and that the Wilson coefficient entering in the operator product expansion, which relates the physical process and the matrix element, can be computed in perturbation theory. The upper bound guarantees small lattice artefacts. In our case, reformulating the inequality in unit of $w_0$ , and setting $w^{\chi}_0 \Lambda \sim w^{\chi}_0 m_V \sim 1$ we have: 
\be
 (w^{\chi}_0 m_V)^2 \sim 1 <  \left(w_0^{\chi}p\right)^2 <  \mcO\left((w_0^{\chi}/a)^2\right)\,.
\ee
Since the smallest value  $w_0^{\chi}/a$ obtained at $\beta=1.8$ is  $w_0^{\chi}/a\sim 2$, we would have $1 <  \left(w_0^{\chi}p\right)^2 <  \mcO\left(4\right)$. This is the famous window problem occurring at coarser lattice spacing. We thus have to  relax the upper bound of the inequality and introduce larger cut-off effects for our coarser lattices. In practice we chose $(w_0^\chi p)^2 = 7$, which corresponds to the lattice cutoff at $\beta=2.0$.

In the following we will check that this particular choice of the reference scale does not affect scale-independent quantities, by using a second reference momentum, at the higher end of the sensible momenta range, namely:  $(w_0^\chi p)^2 = 17$. As shown below, our final results are very stable and do not depend, within errors, on the particular choice of reference momentum.


We summarise the values of the renormalisation constants, defined at our reference scale $(w_0^\chi p)^2 = 7$, for the four $\beta$ values, in \tab{table:Zx}.

\begin{table}[t]
  \begin{tabular}{ccccc}
    \hline
    $\beta$ & $Z_A$   & $Z_V$  & $Z_P/Z_S$  & $Z_P^{\rm{RI'}}(p^2 = 7/w_0^2)$   \\
    \hline
    1.8  & 0.7791(4)(9) & 0.5599(4)(40) & 0.2809(48)(45) & 0.2051(36)(66)        \\
    2.0  & 0.8072(3)(5) & 0.6356(2)(26) & 0.4080(25)(27) & 0.2907(16)(72)        \\
    2.2  & 0.8267(2)(23) & 0.6973(2)(30) & 0.5655(16)(121) & 0.3803(8)(49)      \\
    2.3  & 0.8449(23)(72) & 0.7280(19)(80) & 0.6799(260)(440) & 0.4201(136)(13) \\
    \hline
  \end{tabular}
  \caption{Renormalisation constant obtained using $(w_0 p)^2 = 7$ as a reference scale\label{table:Zx}}
\end{table}


\section{Spectroscopy}\label{sec:spectro}

\subsection{Effective Masses}

\begin{figure}[p!]
\centering
\begin{minipage}{.48\textwidth}
  \centering
  \includegraphics[width=\linewidth]{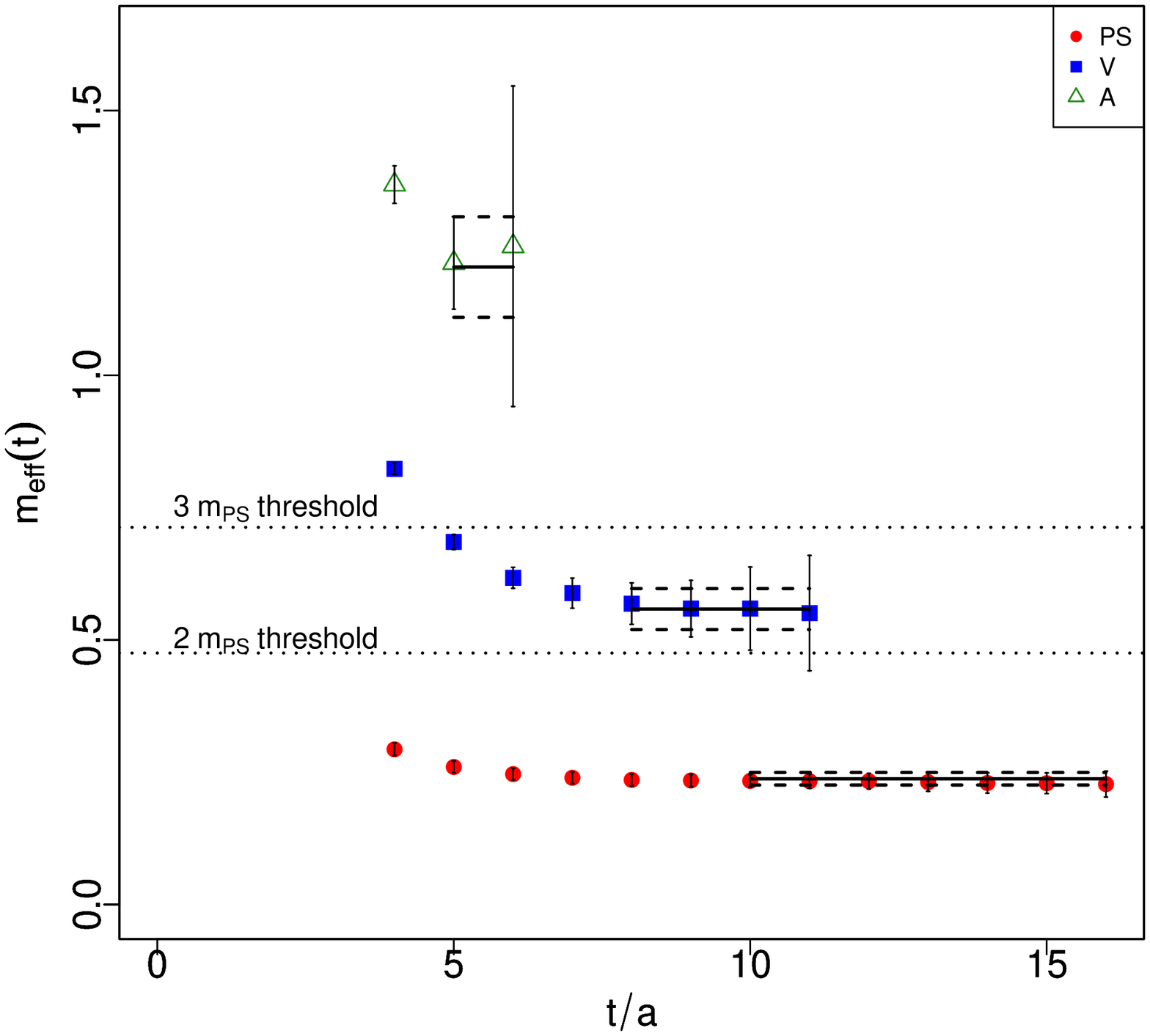}
  \caption{Effective masses of the pseudoscalar, vector and axial meson masses ($\beta=1.8$, $m_0=1.157$, $L=24$)}
  \label{fig:meff_b18_1157}
\end{minipage}%
\hspace*{0.5cm}\begin{minipage}{.48\textwidth}
\centering
   \includegraphics[width=\linewidth]{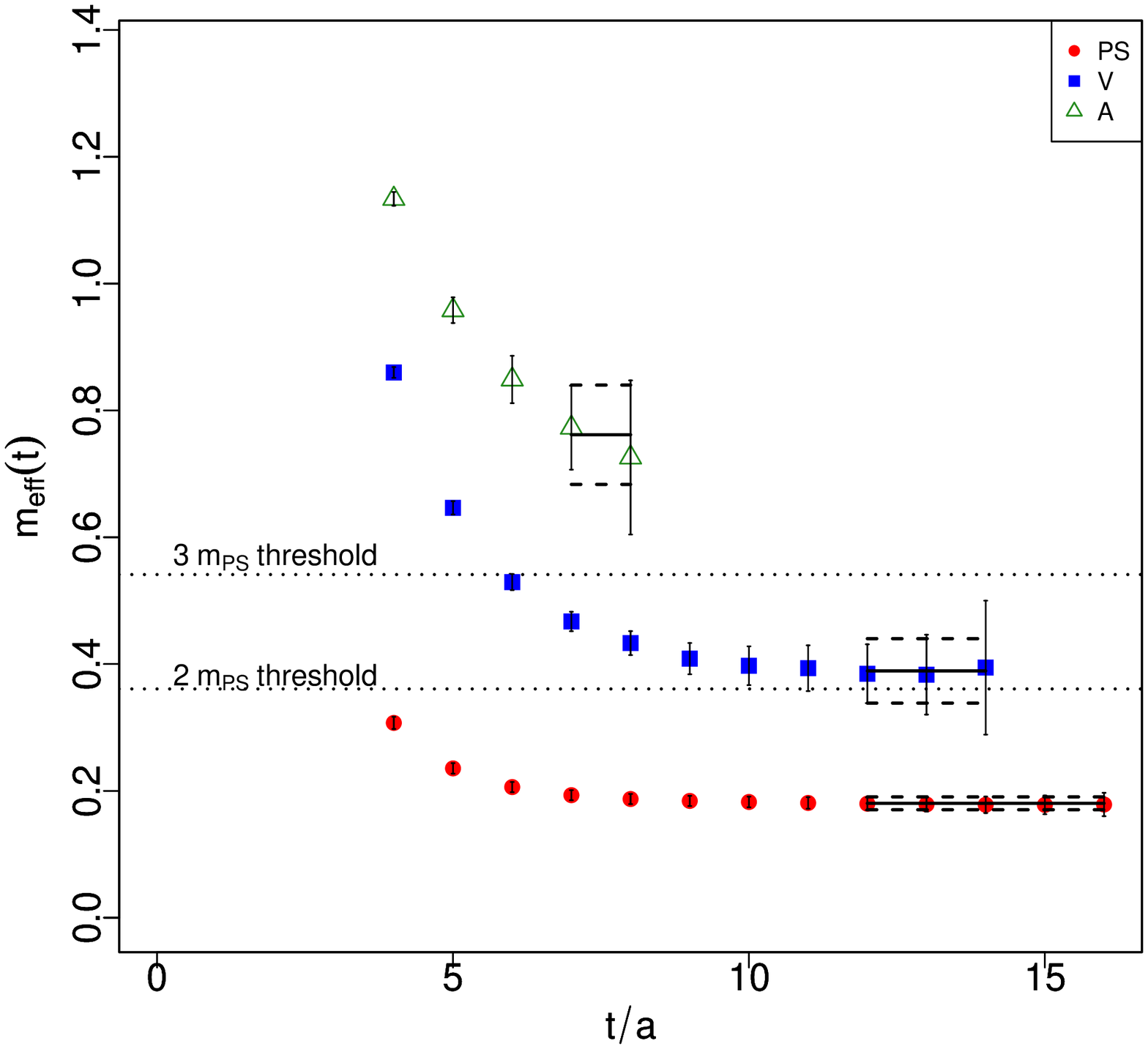}
  \caption{Effective masses of the pseudoscalar, vector and axial meson masses ($\beta=2.0$, $m_0=0.958$, $L=32$)}
  \label{fig:meff_b20_958}
\end{minipage}
\end{figure}
\begin{figure}[p!] 
\centering
\begin{minipage}{.48\textwidth}
  \centering
 \includegraphics[width=\linewidth]{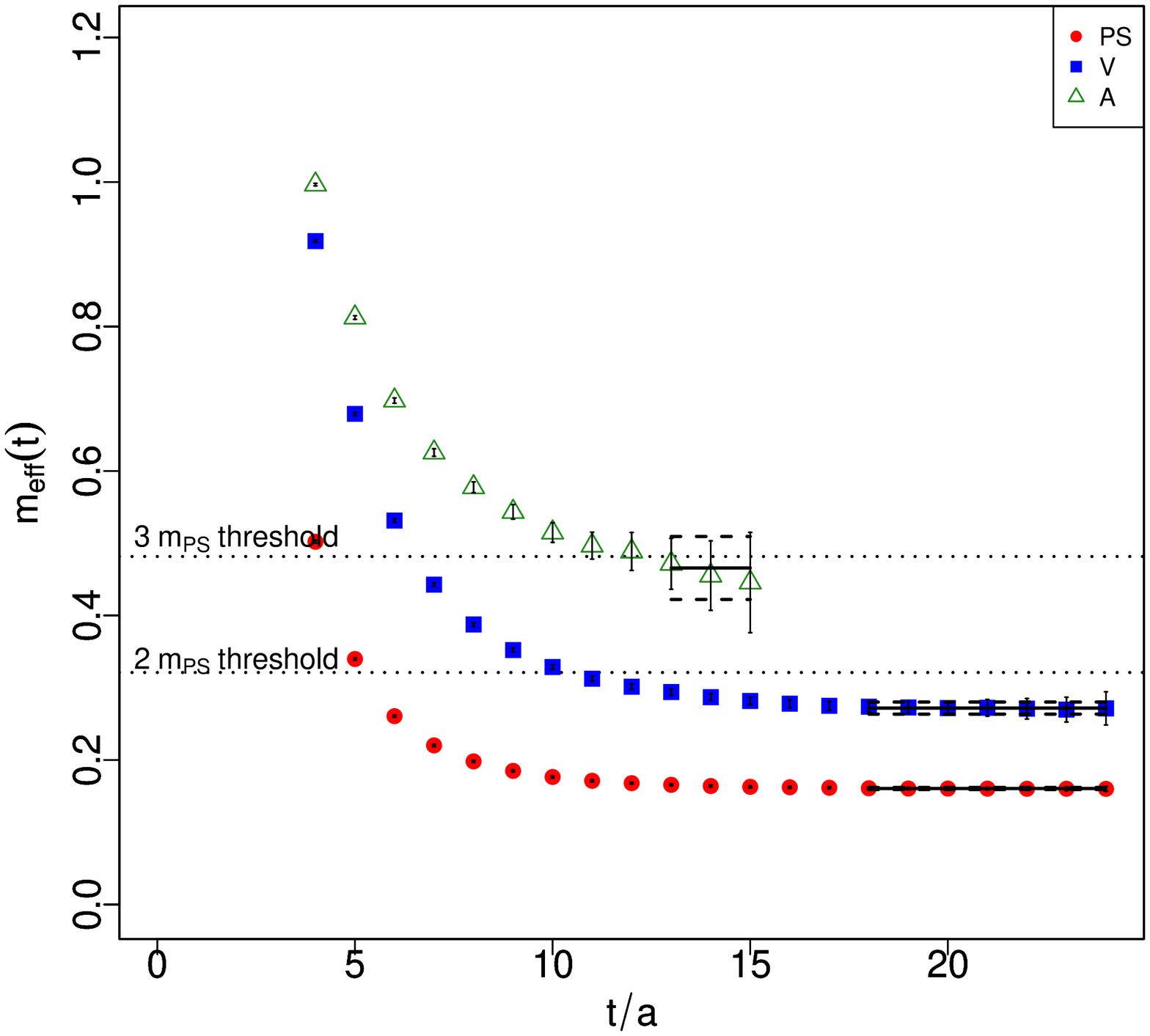}
  \caption{Effective masses of the pseudoscalar, vector and axial meson masses ($\beta=2.2$, $m_0=0.76$, $L=48$)}
  \label{fig:meff_b22_76}
\end{minipage}%
\hspace*{0.5cm}\begin{minipage}{.48\textwidth}
\centering
   \includegraphics[width=\linewidth]{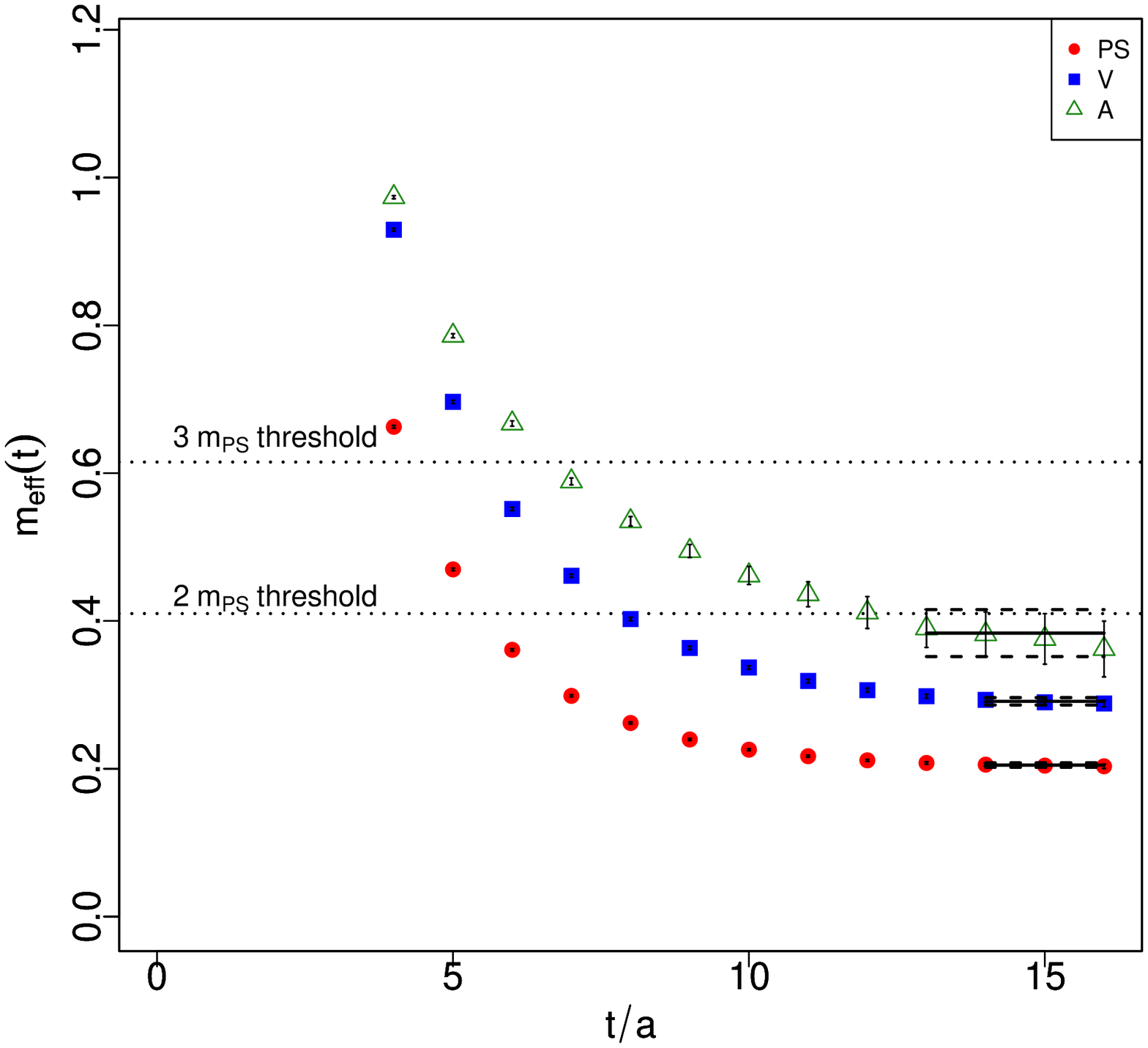}
  \caption{Effective masses of the pseudoscalar, vector and axial meson masses ($\beta=2.3$, $m_0=0.675$,$L=32$)}
  \label{fig:meff_b23_675}
\end{minipage}
\end{figure}

We compute the mass of the lightest (isovector) pseudoscalar, vector and axial-vector meson resonances using two-point correlators. As explained in Section~\ref{sec:setup}, the mass can be extracted using the large time behaviour of the effective mass as decribed by~\eq{eq:meff}. This approach is justified if the state is stable.
We illustrate effective masses for various ensembles in Fig. \ref{fig:meff_b18_1157}, \ref{fig:meff_b20_958}, \ref{fig:meff_b22_76}  and \ref{fig:meff_b23_675}.

The effective masses are fitted on a given plateau range, which is determined for each state by individual inspection. Systematic errors introduced by the choice of the plateaux are small for the pseudoscalar and vector resonances, and for this reason we will neglect them in the following. The best fit value for the effective mass is plotted for each state in the figures together with its statistical error. The masses of the vector and pseudoscalar mesons are clearly determined for all ensembles. For the axial vector  correlator we do not observe long plateaux, due to the much worse signal-to-noise ratio as a function of Euclidean time separation. This results in significantly larger systematic errors, which are not yet fully under control.

In each plot, we also show the two- and three-pion thresholds. This shows that the vector meson resonance, whose main decay channel is expected to be the decay in two pions, is stable for almost all of our simulations. In a few cases, our most chiral points at $\beta=1.8$ and $\beta=2.0$ are at kinematical threshold. A similar conclusion can be drawn for the isovector axial-vector meson, whose main decay channel is expected to be three pions.

\subsection{ $\mps$ and $\fps$}

\begin{figure}[t!]
\centering
\begin{minipage}{.48\textwidth}
  \centering
  \includegraphics[width=\linewidth]{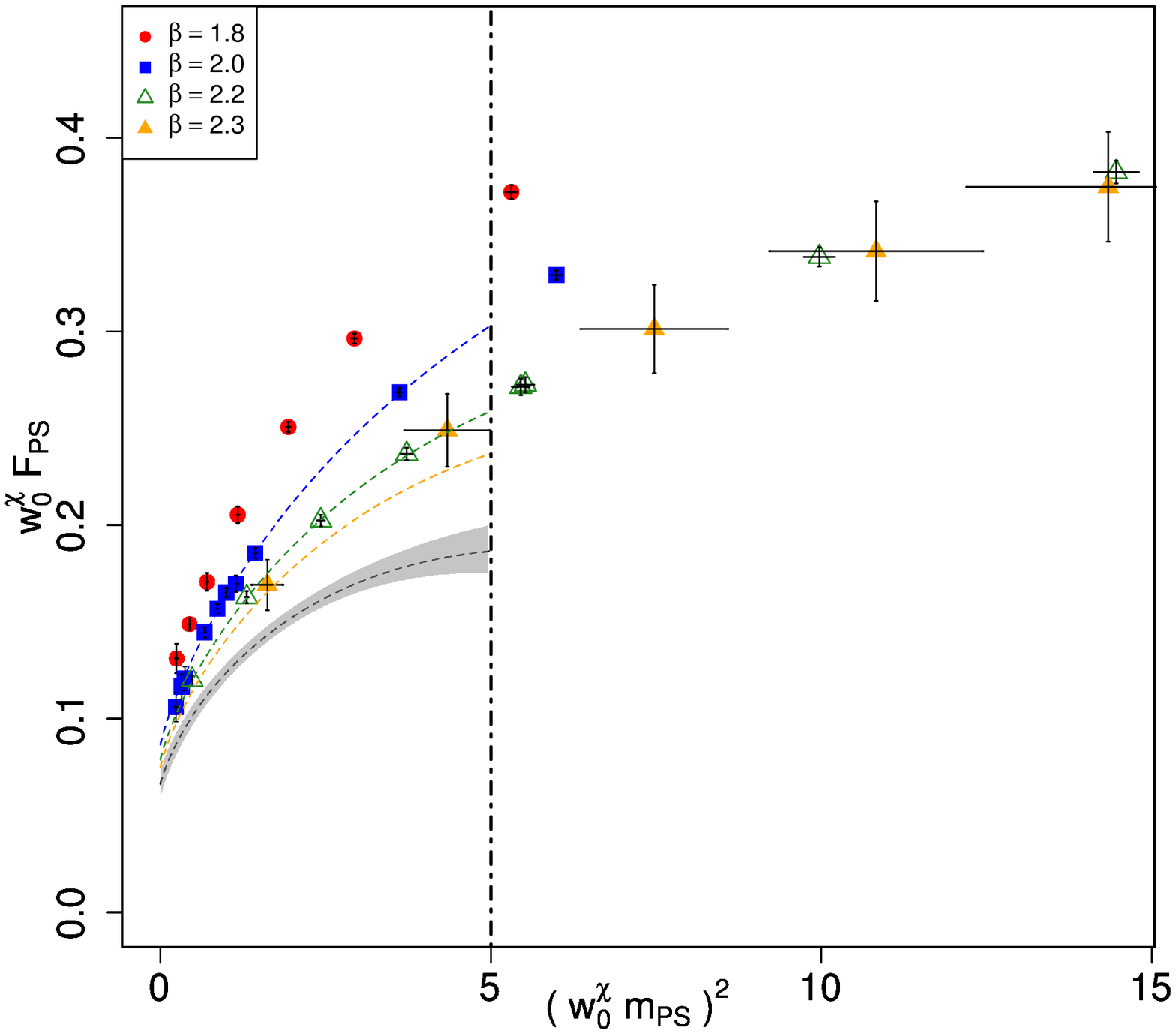}
  \caption{$\fps$ versus $\mps^2$ for the four lattice spacings. The curves correspond to the best fit parameters obtained fitting only $\beta=2.0$, $\beta=2.2$ and $\beta=2.3$ (subset $S_2$) and drawn for the corresponding lattice spacing. The black curve indicate the continuum results.}
  \label{fig:w0fps_vs_mps2}
\end{minipage}%
\hspace*{0.5cm}\begin{minipage}{.48\textwidth}
\centering
  \includegraphics[width=\linewidth]{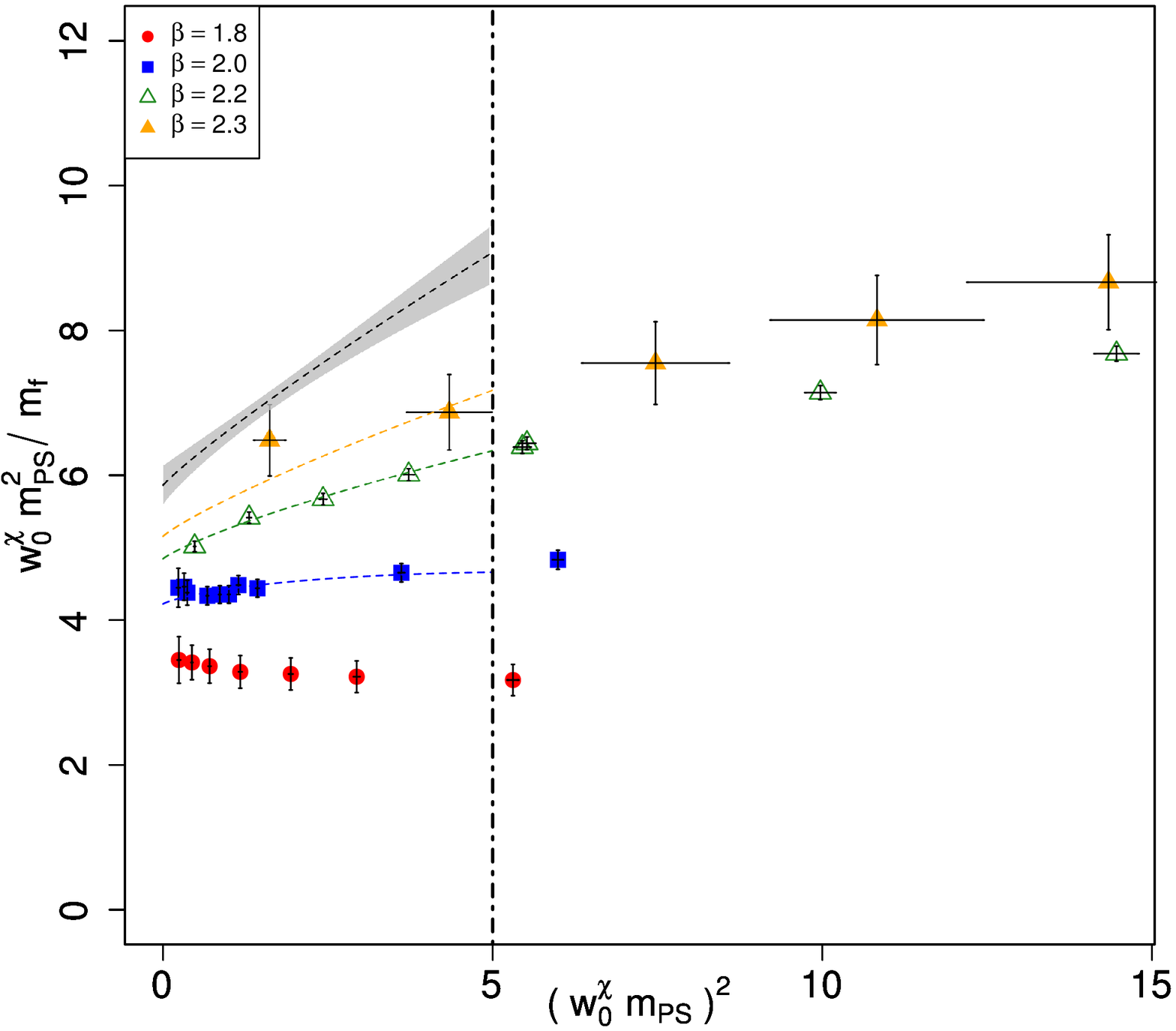}
  \caption{$\mps^2/\mf$ versus $\mps^2$ for the four lattice spacings. The curves correspond to the best fit parameters obtained fitting only $\beta=2.0$, $\beta=2.2$ and $\beta=2.3$ (subset $S_2$) and drawn for the corresponding lattice spacing. The black curve indicate the continuum results.}
  \label{fig:w0mps2_over_mf_vs_mps2}
\end{minipage}
\end{figure}

\begin{table}[p!]
  \begin{tabular}{cccccc}
    \hline\hline
    type & coef.   & $S_1$  & $S_2$  & $S_3$ & $S_4$   \\
    \hline
    NLO global & F             &    0.066(6) & 0.066(6) & 0.049(5) & 0.049(5) \\                                    
    NLO global & $b_F$         &    0.0038(2) & 0.0038(2) & 0.0028(1) & 0.0028(1) \\                                
    NLO global & $\delta_F$    &    0.05(1) & 0.05(1) & 0.09(1) & 0.09(1) \\                                        
    NLO global & $\gamma_F$    &    0.05(1) & 0.051(9) & 0.072(6) & 0.069(6) \\                                     
    NLO global & $a_F$         &    0.22(3) & 0.21(2) & 0.19(1) & 0.18(1) \\                                        
    NLO global & $\chi^2/$ndof &    9.7/8 & 13./10 & 83./14 & 91./16 \\                                             
    NLO global & cut           &    5 & 5 & 5 & 5 \\                                                                                                   
    \hline
    NLO global & B            &  2.9(1) & 2.9(1) & 3.0(1) & 3.0(1) \\                  
    NLO global & $b_M$        &  0.0005(1) & 0.0005(1) & 0.00028(8) & 0.00029(8) \\     
    NLO global & $\delta_M$   &  -0.7(1) & -0.74(9) & -0.84(7) & -0.85(6) \\            
    NLO global & $\gamma_M$   &  -0.25(7) & -0.25(7) & -0.24(5) & -0.24(5) \\           
    NLO global & $a_M$        &  0.00(1) & 0.00(1) & 0.003(5) & 0.003(5) \\             
    NLO global & $\chi^2/$ndof&  10./8 & 14./10 & 27./14 & 30./16 \\                    
    NLO global & cut          &  5 & 5 & 5 & 5 \\                                      
\hline\hline                              
 \end{tabular}
  \caption{Results of the global fits of $\mps^2/\mf$ and $\fps$  on subset $S_{1,2,3,4}$ using $(w_0 p)^2 = 7$ as a reference renormalisation scale. \label{tab:NLO_free}}
\end{table}

\begin{table}[p!]
  \begin{tabular}{cccccc}
    \hline\hline
    type & coef.   & $\beta=1.8$  & $\beta=2.0$  & $\beta=2.2$  & $\beta=2.3$  \\
    \hline
    NLO fixed $\beta$&  F           &   0.096(4) & 0.088(3) & 0.086(3) & 0.09(4) \\                 
    NLO fixed $\beta$&  $a_F$        &  0.41(2) & 0.27(1) & 0.211(9) & 0.1(1) \\                    
    NLO fixed $\beta$ &  $b_F$        &  0.0093(1) & 0.0066(1) & 0.0052(1) & 0.004(1) \\             
    NLO fixed $\beta$ &  $\chi^2/$ndof&  4.1/4 & 9.3/7 & 6.8/4 & 1.1/1 \\                            
    NLO fixed $\beta$ &  cut          &  12 & 12 & 12 & 12 \\                                        
\hline
    NLO fixed $\beta$  &   B              &   1.7(1) & 2.18(8) & 2.38(4) & 3.1(7) \\               
    NLO fixed $\beta$&   $a_M$          &   -0.02(6) & -0.00(2) & 0.025(7) & 0.01(9) \\          
    NLO fixed $\beta$&   $b_M$          &   -0.0004(7) & 0.0000(3) & 0.0007(1) & 0.000(1) \\     
    NLO fixed $\beta$&   $\chi^2/$ndof  &   4.0/4 & 8.2/7 & 5.0/4 & 1.0/1 \\                     
    NLO fixed $\beta$  &   cut            &   12 & 12 & 12 & 12 \\                                 
\hline\hline
  \end{tabular}
  \caption{Results of the fixed lattice spacing fits for each $\beta$ value using $(w_0 p)^2 = 7$ as a reference renormalisation scale. \label{tab:NLO_fixed}}
\end{table}

The continuum expressions for $\mps$ and $\fps$ have been worked out in~\cite{Bijnens:2009qm} at next-to-leading order in chiral perturbation theory:
\beq\label{eq:NLO_cont_mps_fps}
\frac{\mps^2}{\mf} &=& 2 B \left[ 1 +\frac{3}{4} x \log{\frac{2 B \mf}{\mu^2}} +   b_M  x +\mcO(x^2)  \right]\,,\\
\label{eq:NLO_cont_mps_fps2}
\fps &=&\phantom{2} F\left[ 1 - x\log{\frac{2 B \mf}{\mu^2}} +  b_F x +\mcO(x^2) \right]\,,
\eeq
where $x = \frac{2 B \mf}{(4\pi F)^2}$ and $\mf$ is the renormalised
fermion mass at a given scale. In the conventions of \cite{Bijnens:2009qm}, the condensate is given by $\Sigma \equiv - 2 B F^2$. 
Note that $F$ and $B$ appear in both expressions. 
The range of applicability of the effective theory is not known \textit{a priori}. 
In order to make the fits more stable, we will rewrite the expansion in a new parameter, $\tilde{x} = \frac{\mps^2}{(4\pi F)^2}$. At this order \eq{eq:NLO_cont_mps_fps} and \eq{eq:NLO_cont_mps_fps2} remain unchanged (this is, however, not true at NNLO) and read:
\beq\label{eq:NLO_cont_mps_fps_xtilde}
\frac{\mps^2}{\mf} &=& 2 B  \left[ 1 +\frac{3}{4} \tilde{x} \log{\frac{\mps^2}{\mu^2}} +   b_M  \tilde{x} +\mcO(\tilde{x}^2) \right]\,,\\
\fps &=& \phantom{2}F\left[ 1 - \tilde{x}\log{\frac{\mps^2}{\mu^2}} +  b_F \tilde{x}  +\mcO(\tilde{x}^2) \right]\,.
\eeq
From this result we observe that the expansion of $\fps$ now is independent of $B$, which will allow us to perform the fit in two steps: first a fit to $\fps$ to obtain $F$ and then using it as an input for a second fit to $\mps^2/\mf$ to obtain $B$.

The renormalised values for  $\fps$ and $\mps^2/\mf$ at four values of the lattice spacing are shown as function of $\mps^2$ in \fig{fig:w0fps_vs_mps2} and \ref{fig:w0mps2_over_mf_vs_mps2}. 
All the lattices included in the fit satisfy $\mps L \ge 5.6$. 
The fermion mass is given by $\mf(p^2)= \mpcac\,Z_A/Z_P(p^2)$ and the renormalised pseudoscalar decay constant is $\fps =\fps^{\rm (bare)} Z_A$. As a reference scale for the renormalisation constants we use $p=\sqrt{\,7\,}/w^\chi_0$.
As can be seen, significant cut-off effects are observed. 
In order to estimate the low energy constants $F$ and $B$ in the continuum, discretisation effects must then be taken into account. In order to obtain a reliable estimate, we will use two different strategies.

The first strategy (strategy I) is based on fitting the pseudoscalar mass and decay constant using several lattice spacings simultaneously together with a given model for the lattice discretisation effects:
\beq\label{eq:NLO_mps_fps}
\frac{\mps^2}{\mf} &=& 2 B  \left[ 1 - a_M \tilde{x} \log{\frac{\mps^2}{\mu^2}} +   b_M  \tilde{x}   +\delta_M \frac{a}{w^{\chi}_0} +\gamma_M \mps^2 \frac{a}{w^{\chi}_0}  \right]\,, \\
\fps &=& F\left[ 1 -a_F \tilde{x}\log{\frac{\mps^2}{\mu^2}} +  b_F \tilde{x}  +\delta_F \frac{a}{w^{\chi}_0} +\gamma_F \mps^2 \frac{a}{w^{\chi}_0}  \right]\label{eq:NLO_mps_fps2}\,.
\eeq
Here the new fitting parameters $\delta_{M,F}$ and $\gamma_{M,F}$ control the discretisation effects. Note that the two coefficients $a_{F,M}$ are fixed in the continuum, but here we consider them as free parameters.

To control the stability of the fit, we consider four subsets of our data $S_1=\{\beta=2.0, 2.2\}$, $S_2=\{\beta=2.0, 2.2, 2.3\}$, $S_3=\{\beta=1.8, 2.0, 2.2\}$ and $S_4=\{\beta=1.8, 2.0, 2.2, 2.3\}$ and perform the fit on each of these subsets.
The result of the fit for the $S_2$ subset is shown in \fig{fig:w0fps_vs_mps2} and \ref{fig:w0mps2_over_mf_vs_mps2}.

The second strategy (strategy II) consists of fitting each of the lattice spacings independently, to obtain the coefficients $B,F,a_{F,M}$ and $b_{F,M}$, while setting to zero the coefficients $\delta_{M,F},\gamma_{M,F}$ in Eq. \eqref{eq:NLO_mps_fps} and (\ref{eq:NLO_mps_fps2}). In a second step, lattice discretisation effects can be assessed by studying the dependence of the coefficients as a function of the lattice spacing.

In all fits we use $w_0^{\chi }\mu=1$ as a scale. 
The results of the fits, including their $\chi^2$ per degrees of freedom, are summarised in Table~\ref{tab:NLO_free} for strategy I and Table~\ref{tab:NLO_fixed} for strategy II. The fits are performed on a given range of values for $(w_0^{\chi} \mps)^2$ below the ``cut'' given 
 in the tables.  

\begin{figure}[t!] 
\includegraphics[width=0.5\textwidth]{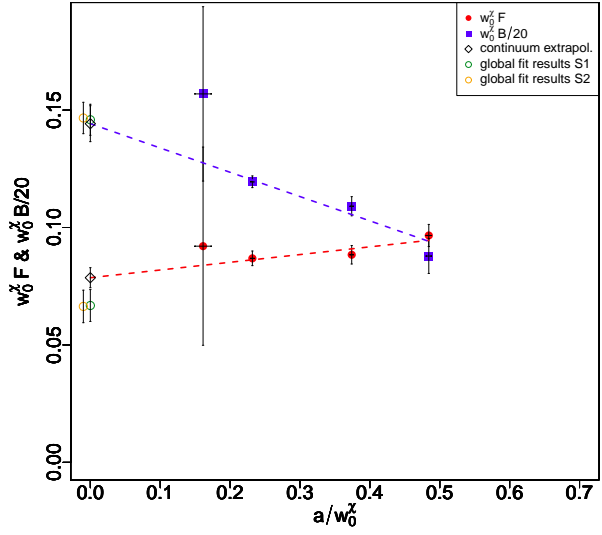}
\caption{Values of the chiral parameters $B$ and $F$ in units of the reference lattice scale $w_0^\chi$ as extracted using strategy II described in the text. In this plot $B$ has been rescaled by a factor of 20 for graphical convenience. \label{fig:scaling}}
\end{figure}

Strategy II allows us to extract an estimate of $w_0^{\chi} F$ and $w_0^{\chi} B$ for each lattice spacing. This is shown in \fig{fig:scaling}, where the value of $B$ has been re-scaled by a factor of 20 for convenience.
The scaling towards the continuum limit is compatible with a linear behaviour and no $\mcO(a^2)$ effects are visible. 
On the plot we also show the results obtained directly in the continuum using the first strategy for the subset of gauge ensembles $S_1$ and $S_2$. 
The results obtained with strategy I for the subsets $S_3$ and $S_4$ have a $\chi^2/\textrm{ndof} \sim 10$ and thus do not describe the data well.

Our final estimates for the chiral parameters are $w_0^{\chi} B=2.88(15)(17)$ and $w_0^{\chi} F=0.078(4)(12)$. The central value and statistical error comes from the linear extrapolation to the continuum  of the fits at fixed beta (strategy I). The systematic error is obtained by computing the maximal difference between the results obtained by strategy I and II. By setting the scale to $F=246~\GeV$ one arrives at the result $w_0^{\chi} =6.3 (3)(9)\cdot 10^{-5}~\fm$. 
The value of the condensate then reads $\Sigma^{1/3}/F = 4.19(26)$ (statistical and systematical errors have been combined).

We repeated a similar analysis using $p = \sqrt{17}/w^\chi_0 $ as reference scale, which is shown in appendix \ref{ap:NPR}. As claimed in the previous section, we do not observe any statistically significant change in the continuum values of $F$ and $B$.

\begin{figure}[p!] 
\includegraphics[width=0.5\textwidth]{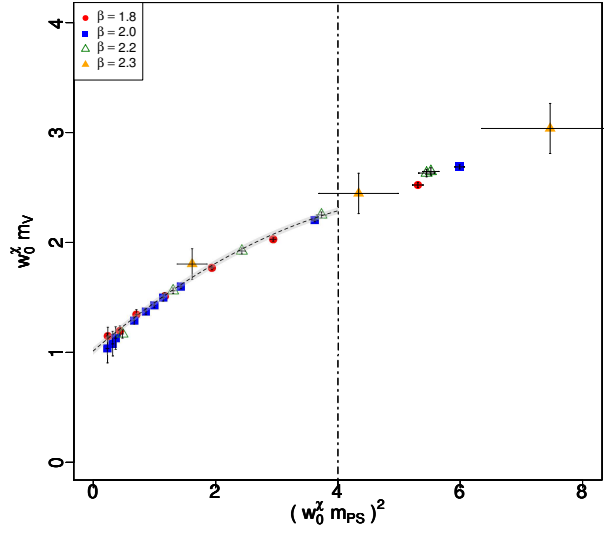}
\caption{Combined chiral and continuum extrapolation of the vector meson mass $m_V$. Our data for four lattice spacings is presented together with the best fit at each lattice spacing. The grey band is our result for the continuum extrapolation and its 1-$\sigma$ confidence region.\label{fig:w0mV}}
\end{figure}
\begin{figure}[p!] 
\includegraphics[width=0.5\textwidth]{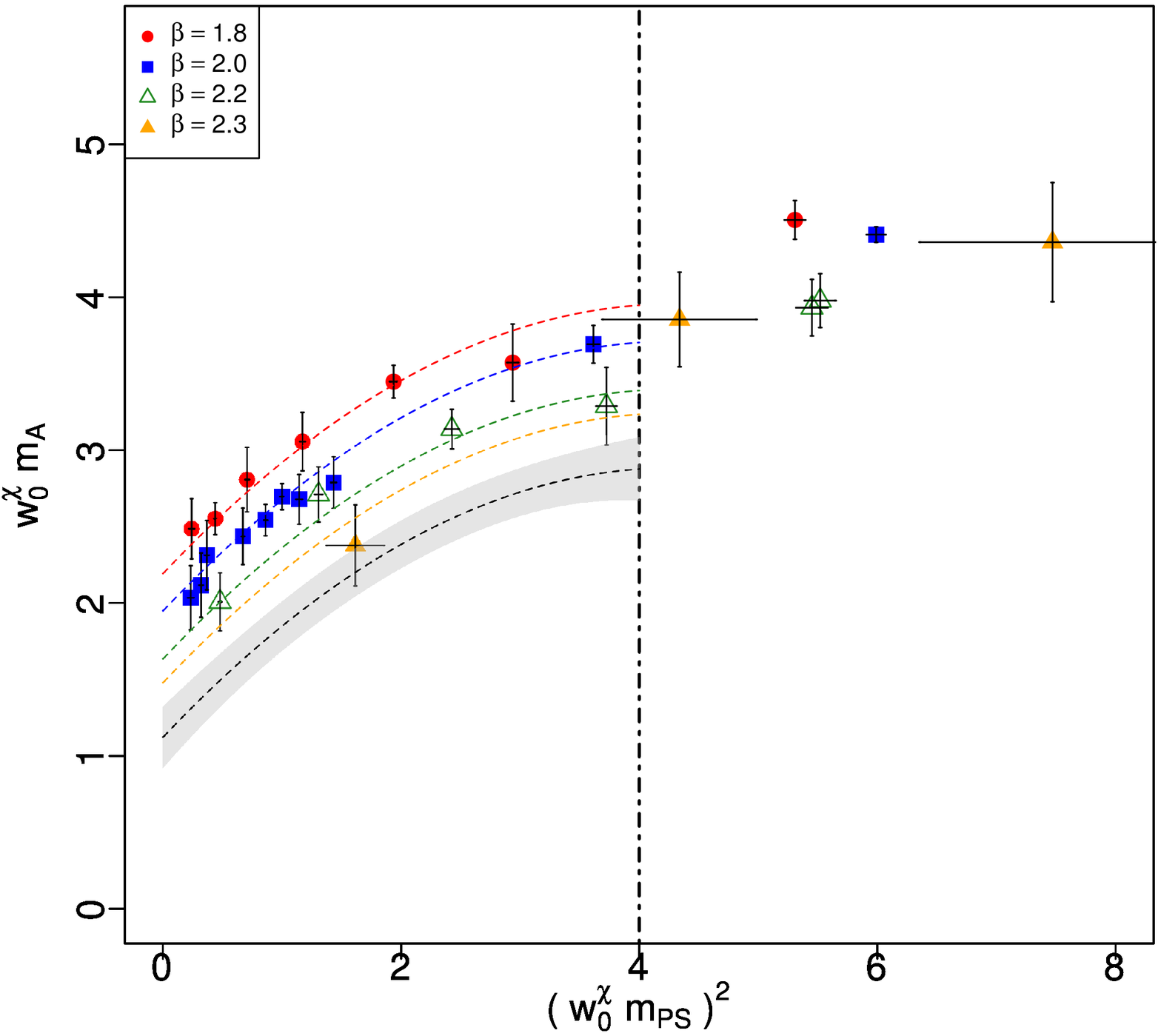}
\caption{Combined chiral and continuum extrapolation of the axial vector meson masss $m_A$. Our data for four lattice spacings is presented together with the best fit at each lattice spacing. The grey band is our result for the continuum extrapolation and its 1-$\sigma$ confidence region.\label{fig:w0mA}}
\end{figure}

\subsection{Heavier states}
 
In this section we report our results for the mass of two heavier isotriplet meson resonances, namely the vector in \fig{fig:w0mV} and the axial-vector in \fig{fig:w0mA}.
All the masses are presented in units of $w_0^\chi$ as functions of $(w^\chi_0 \mps)^2$. 
In each figure we present a global fit, including all the available data at four lattice spacings, to the following fit ansatz:
\be
w_0^\chi m_X = w_0^\chi m_X^\chi + A  (w_0^\chi \mps)^2 + B  (w_0^\chi
\mps)^4 + C \frac{a}{w_0}\, .\label{eq:heavyans}
\ee
The fit range for each channel is shown by the vertical dotted line in the plot.
The gray band indicate the $1\sigma$ error band for the continuum prediction, obtained by setting $a=0$ with our best fit parameters. The results of the fit for the axial and vector meson are summarised in \tab{tab:mV_mA}.

\begin{table}[t!]
  \begin{tabular}{ccc}
    \hline\hline
    coef. & Vector   & Axial   \\
    \hline
    $w^{\chi}_0 m_X$ & 1.01(3)        &     1.1(1)             \\
    $A$            &   0.47(3)    &         0.8(1)        \\
    $B$            &   -0.039(6)  &         -0.09(3)     \\
    $C$            &   -0.05(7)   &         2.1(3)        \\
    $\chi^2/$ndof  &  23/16    & 20/16  \\     
    cut            &  4        & 4 \\
    \hline\hline
  \end{tabular}
  \caption{Results of the polynomial fits of the vector and axial resonances. \label{tab:mV_mA}}
\end{table}

For the vector meson the fit describes our data well and the observed cutoff effects are small. We find $w_0^{\chi} m^\chi_V = 1.01(3)$ with a $\chi^2/\textrm{ndof}=23/16$. Note that for our data $m_V$ is always less than $2\mps$, except maybe for the most chiral point used in the fit, so that the vector meson is expected to be stable and its mass can be reliably extracted from the large (Euclidean) time behaviour of the appropriate two-point function.

For the mass of the axial-vector meson, our data is more noisy already at the level of the effective masses and we therefore have larger systematic uncertainties.  The ansatz \eq{eq:heavyans} fits the data well, within large errors, and the resulting value for the mass is: $w_0^\chi m^\chi_A =1.1(1)$ with $\chi^2/\textrm{ndof}=20/16$.
In units of $\fps$ we have $m_V/\fps\sim 13.1(2.2)$ and $m_A/\fps\sim 14.5(3.6) $.


\section{Conclusion}

We analysed the  SU(2) gauge theory with $N_f = 2$ flavours of fermions in the
fundamental representation using lattice techniques. Dynamical simulations have been performed at four lattice spacings and a number of volumes and masses to asses systematic effects and to carry out the necessary extrapolations. We determined non-perturbatively, in the RI'-MOM scheme, the relevant renormalisation constants and performed a detailed analysis of the mass and decay constant of the pseudoscalar Goldstone bosons, including an extrapolation to the chiral and continuum limits to take into account the lattice cutoff effects present in our computation. We use a conservative estimate of all systematic uncertainties to obtain a reliable estimate of $\fps$. Finally we analysed the mass of the spin-1 bound states and determined the ratios $m_V/\fps=13.1(2.2)$ and $m_A/\fps=14.5(3.6)$ for the continuum theory in the chiral limit, using similar extrapolation methods. Our final results are consistent with, and improve upon, previous results for this model, which were performed with only two lattice spacings, at much larger quark masses and using a perturbative estimate of the renormalisation constants.

In the context of the fundamental composite (Goldstone) Higgs dynamics \cite{Cacciapaglia:2014uja} our results predicts new resonances of mass:
\be
m_V = \frac{3.2(5)}{\sin \theta}~\TeV,\quad\text{and}\quad m_A = \frac{3.6(9)}{\sin \theta}~\TeV\,,
\ee
which are beyond the present LHC constraints, even in the Technicolor limit \cite{Franzosi:2015zra} where $\theta=\pi/2$.

In the context of dark matter models, in paticular for the SIMPlest case, because the dark pion is estimated to be around ten times its decay constant \cite{Hansen:2015yaa}, we cannot use the estimate above. Nonetheless, a preliminary result can be obtained from our simulations reported in the first line of Table~\ref{ap:res2}, at $\beta =2$, which yield $m_{\rm PS}/\fps \approx 7.5 $,  $m_V/\fps \approx 8.3 $ and  $m_A/\fps \approx 13.7 $. Although these results need crucial refinement they immediately show that for such large values of the dark pion mass one cannot neglect the effects of higher mass states since the overall spectrum is much more compressed than in the case of the chiral limit.

\section*{Acknowledgments}
This work was supported by the Danish National Research Foundation DNRF:90 grant and by a Lundbeck Foundation Fellowship grant. We acknowledge PRACE for awarding us access to computational resources on MareNostrum at the Barcelona Supercomputing Centre, Spain.
Additional local computational facilities used in this work were provided by the local HS9 cluster and by the DeIC national HPC centre at SDU.


\clearpage
\begin{appendix}

\section{Numerical results}
We report in this section our numerical results for the main spectroscopy quantities studied in this article.  The column ``stat'' reports the number of thermalised configurations used in the analysis, while the column ``$N_{\rm{rep}}$'' is the number of ``replicas'' runs used, i.e. number of independent runs with the same bare parameters.
\bigskip

\label{ap:res}
\begin{longtable}{cccccccccccc}
    \hline
    $\beta$ & L & T & $m_0$  & $N_{\rm{rep}}$ & stat. & $\mps L$ & $\mpcac^{(bare)}$ & $\mps$ & $\fps^{(bare)}$ &  $m_V$ & $m_A$ \\
    \hline
    \hline
    \endhead
\hline \multicolumn{12}{c}{{Table continued on next page}} \\ 
\endfoot
\endlastfoot
1.8 & 16 & 32 & 1 & 1 & 1562 & 17.85136 & 0.2133(2) & 1.115(1) & 0.231(1) & 1.221(1) & 2.20(3)\\
1.8 & 16 & 32 & 1.089 & 2 & 19986 & 13.27994 & 0.11638(7) & 0.8299(3) & 0.1842(3) & 0.9831(9) & 1.7(1)\\
1.8 & 16 & 32 & 1.12 & 1 & 3168 & 10.78498 & 0.0758(2) & 0.674(1) & 0.155(1) & 0.857(4) & 1.6(2)\\
1.8 & 16 & 32 & 1.14 & 1 & 1225 & 8.386688 & 0.0454(5) & 0.524(3) & 0.127(2) & 0.73(1) & 1.5(1)\\
1.8 & 16 & 32 & 1.15 & 1 & 1517 & 6.51176 & 0.0267(5) & 0.406(4) & 0.106(2) & 0.65(2) & 1.3(1)\\
1.8 & 24 & 32 & 1.155 & 1 & 3316 & 7.696368 & 0.0163(3) & 0.320(3) & 0.092(1) & 0.58(2) & 1.2(2)\\
1.8 & 24 & 32 & 1.157 & 1 & 1447 & 5.70156 & 0.0088(7) & 0.23(1) & 0.081(4) & 0.55(3) & 1.34(9)\\
\hline
2 & 16 & 32 & 0.85 & 2 & 46057 & 14.64522 & 0.1669(1) & 0.9153(5) & 0.1524(3) & 1.0050(8) & 1.64(3)\\
2 & 16 & 32 & 0.9 & 2 & 20316 & 11.37803 & 0.1046(2) & 0.711(1) & 0.1244(5) & 0.824(1) & 1.39(5)\\
2 & 16 & 32 & 0.94 & 2 & 9377 & 7.160768 & 0.0434(3) & 0.447(2) & 0.086(1) & 0.598(6) & 1.07(7)\\
2 & 16 & 32 & 0.945 & 1 & 3760 & 6.399184 & 0.0343(6) & 0.399(4) & 0.078(1) & 0.56(1) & 1.07(7)\\
2 & 32 & 32 & 0.947 & 2 & 1826 & 11.96282 & 0.0309(3) & 0.373(2) & 0.0765(9) & 0.535(7) & 1.01(7)\\
2 & 32 & 32 & 0.949 & 4 & 1633 & 11.10832 & 0.0266(3) & 0.347(2) & 0.072(1) & 0.51(1) & 0.96(7)\\
2 & 32 & 32 & 0.952 & 1 & 2005 & 9.80432 & 0.0208(3) & 0.306(3) & 0.067(1) & 0.48(1) & 0.94(8)\\
2 & 32 & 32 & 0.957 & 1 & 711 & 6.762944 & 0.0096(5) & 0.211(6) & 0.054(2) & 0.40(4) & 0.95(5)\\
2 & 32 & 32 & 0.958 & 1 & 957 & 5.772608 & 0.0070(6) & 0.18(1) & 0.049(3) & 0.38(5) & 0.84(9)\\
\hline
2.2 & 16 & 32 & 0.6 & 1 & 256 & 14.10846 & 0.2008(3) & 0.881(1) & 0.107(1) & 0.925(2) & 1.31(2)\\
2.2 & 16 & 32 & 0.65 & 1 & 512 & 11.71451 & 0.1489(3) & 0.732(1) & 0.0949(7) & 0.787(2) & 1.17(4)\\
2.2 & 16 & 32 & 0.68 & 1 & 2894 & 8.716016 & 0.0914(2) & 0.544(1) & 0.0764(6) & 0.613(2) & 0.92(3)\\
2.2 & 16 & 32 & 0.7 & 1 & 2148 & 8.660944 & 0.0909(3) & 0.541(2) & 0.0760(7) & 0.610(3) & 0.91(4)\\
2.2 & 32 & 32 & 0.72 & 2 & 4437 & 14.31942 & 0.0660(4) & 0.4474(7) & 0.0663(4) & 0.521(1) & 0.81(1)\\
2.2 & 32 & 32 & 0.735 & 4 & 1257 & 11.55728 & 0.0456(2) & 0.361(1) & 0.0567(4) & 0.446(3) & 0.72(3)\\
2.2 & 32 & 32 & 0.75 & 5 & 196 & 8.477056 & 0.0257(4) & 0.264(1) & 0.0456(6) & 0.362(4) & 0.62(3)\\
2.2 & 48 & 48 & 0.76 & 1 & 1409 & 7.70736 & 0.0101(1) & 0.160(1) & 0.0337(4) & 0.271(8) & 0.45(7)\\
\hline
2.3 & 32 & 32 & 0.575 & 2 & 717 & 19.54832 & 0.1327(2) & 0.610(1) & 0.0715(5) & 0.648(2) & 0.89(2)\\
2.3 & 32 & 32 & 0.6 & 2 & 4750 & 16.98768 & 0.1066(1) & 0.5308(6) & 0.0651(1) & 0.5731(9) & 0.82(1)\\
2.3 & 32 & 32 & 0.625 & 2 & 1233 & 14.10864 & 0.0793(2) & 0.440(1) & 0.0575(4) & 0.489(2) & 0.72(1)\\
2.3 & 32 & 32 & 0.65 & 2 & 2296 & 10.75213 & 0.0506(2) & 0.336(1) & 0.0475(2) & 0.394(2) & 0.62(1)\\
2.3 & 32 & 32 & 0.675 & 2 & 1436 & 6.560576 & 0.0199(3) & 0.205(3) & 0.0323(5) & 0.291(4) & 0.38(5)\\
  \hline
%
 \caption{Numerical results for large volume runs used in the analysis presented in this paper.\label{table:results}}
 \label{ap:res2}
\end{longtable}

\newpage
\begin{figure}[b]
\centering
\begin{minipage}{.5\textwidth}
  \centering
  \includegraphics[width=\linewidth]{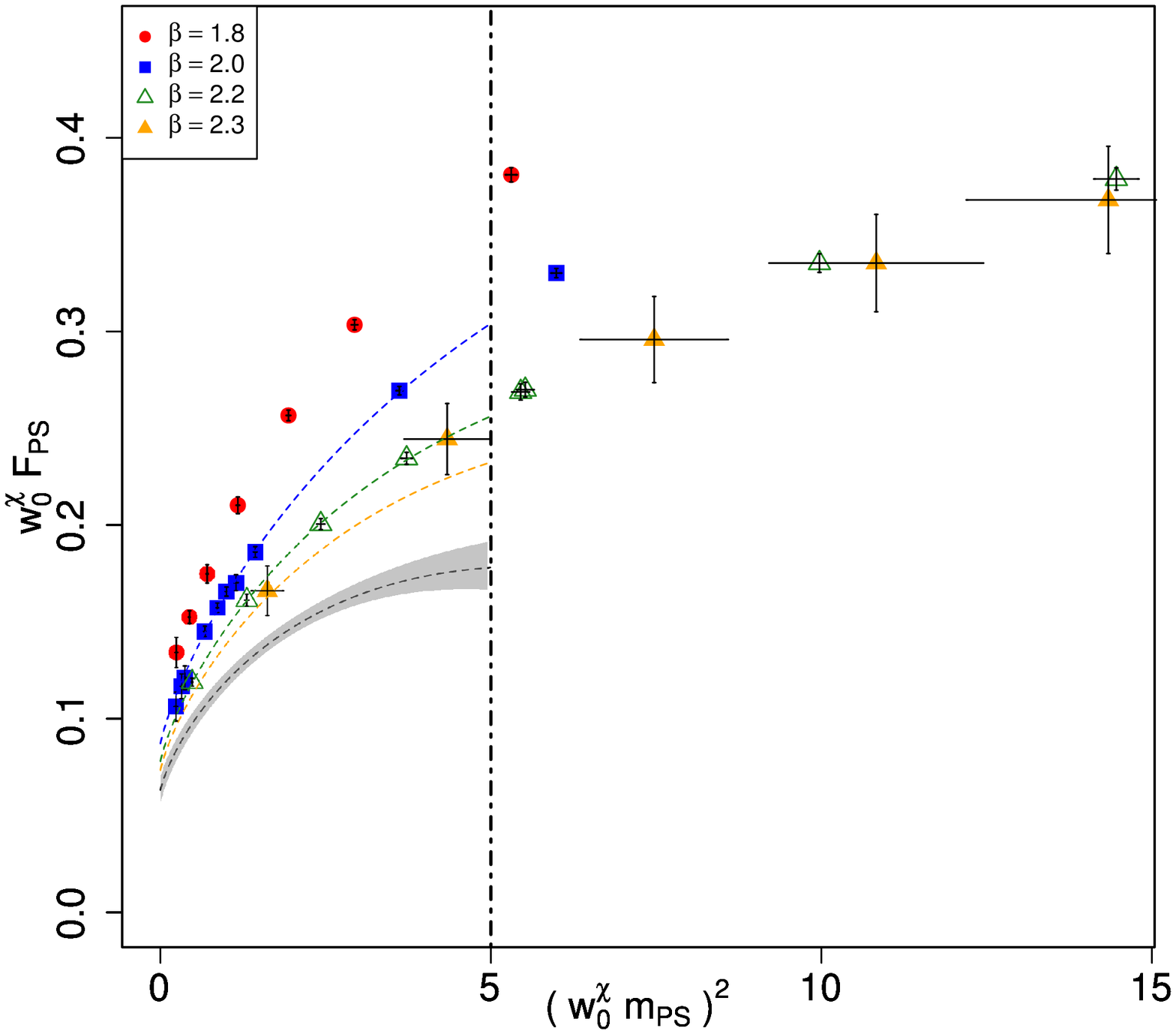}
  \caption{$\fps$ versus $\mps^2$ for four lattice spacing, using $(w_0^\chi p)^2 = 17$ for the renormalisation scale. The curves correspond to the best fit parameters obtained fitting only $\beta=2.0$, $\beta=2.2$ and $\beta=2.3$ (subset $S_2$) and drawn for the corresponding lattice spacing. The black curve indicate the continuum results.}
  \label{fig:w0fps_vs_mps2_17}
\end{minipage}%
\hspace*{0.5cm}\begin{minipage}{.5\textwidth}
\centering
  \includegraphics[width=\linewidth]{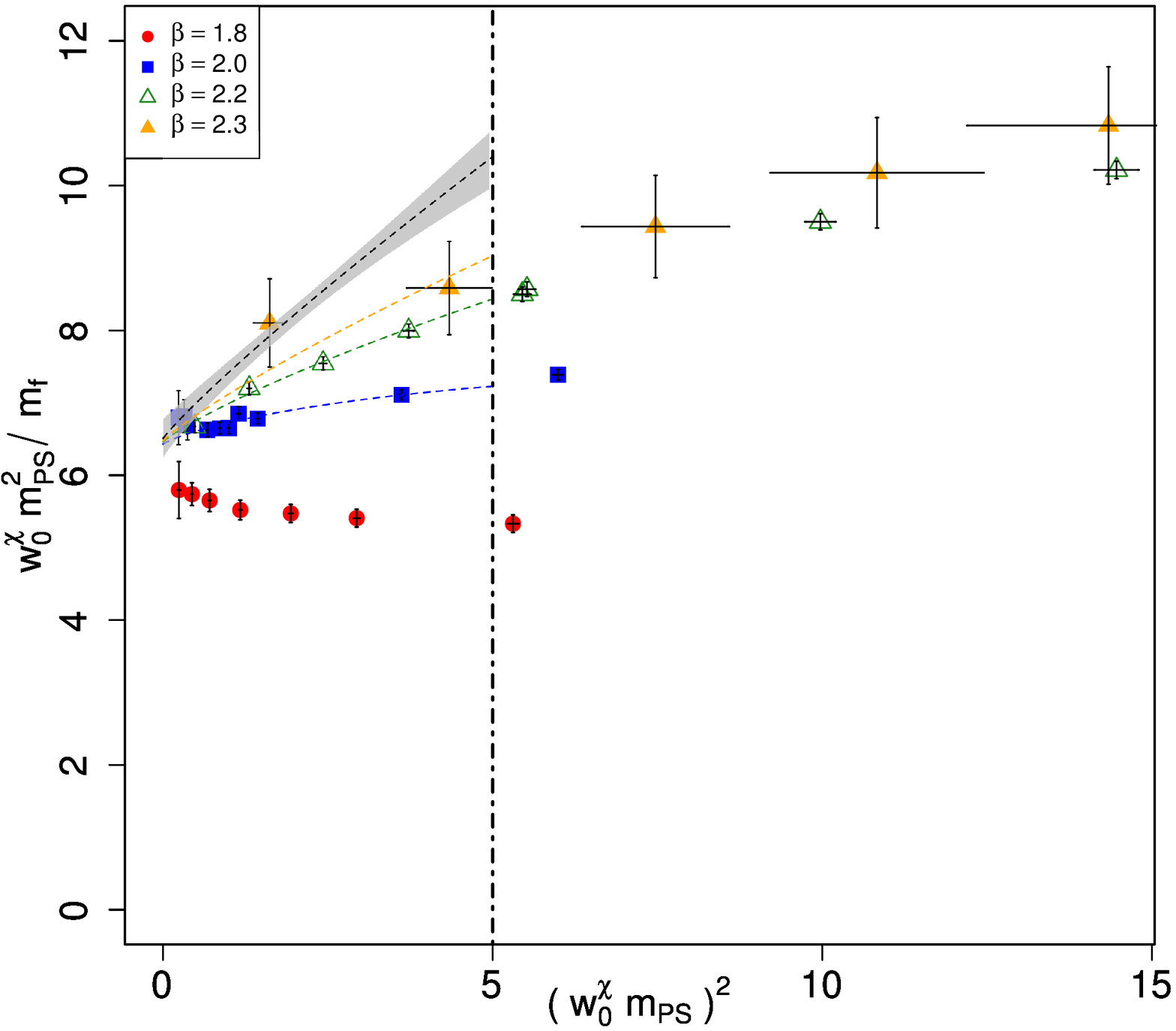}
  \caption{$\mps^2/\mf$ versus $\mps^2$ for four lattice spacing, using $(w_0^\chi p)^2 = 17$ for the renormalisation scale. The curves correspond to the best fit parameters obtained fitting only $\beta=2.0$, $\beta=2.2$ and $\beta=2.3$ (subset $S_2$) and drawn for the corresponding lattice spacing. The black curve indicate the continuum results.}
  \label{fig:w0mps2_over_mf_vs_mps2_17}
\end{minipage}
\end{figure}

\begin{figure}[h!] 
\includegraphics[width=0.5\textwidth]{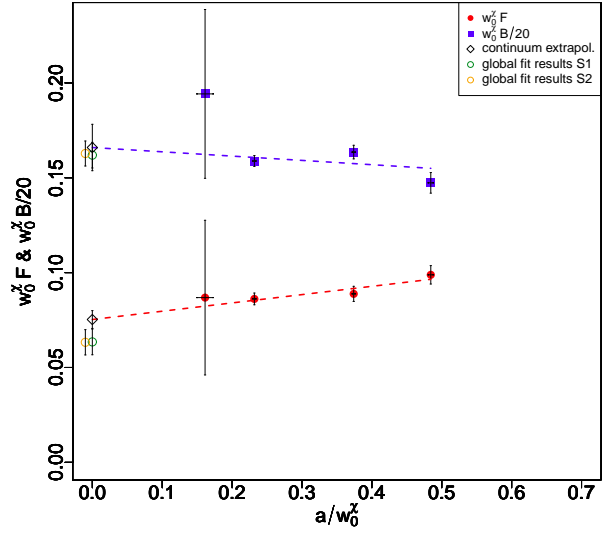}
\caption{Analogue of Fig.~\ref{fig:scaling} obtained for reference momentum $(w_0^\chi p)^2 = 17$ \label{fig:scaling_17}.}
\end{figure}

\section{Systematic error due to the choice of renormalisation scale}
\label{ap:NPR}
In this appendix we report the dependence of our continuum extrapolation results for the low energy constants $F$ and $B$ on the choice of renormalisation scale $(w_0^\chi p)^2$. The main result in the text are obtained using $(w_0^\chi p)^2 = 7$ as the reference momentum scale. Here we present the same analysis for another value of the reference momentum scale: $(w_0^\chi p)^2 = 17$. This corresponds to a much higher scale where lattice cutoff effects are expected to become more relevant. We show below in Figs~\ref{fig:w0fps_vs_mps2_17},~\ref{fig:w0mps2_over_mf_vs_mps2_17} and~\ref{fig:scaling_17} the analysis of $\mps$, $fps$ and the scaling plot $F$ and $B$ using $(w_0^\chi p)^2 = 17$. The corresponding results for the chiral parameters read $w_0^{\chi} B=3.32(24)(8)$ and $w_0^{\chi} F=0.075(5)(12)$. Setting the scale to be $F=246~\GeV$, one thus deduce $w_0^{\chi} =6.0 (4)(9)\cdot 10^{-5}~\fm$.  The value of the condensate then read $\Sigma^{1/3}/F = 4.48(28)$ (statistical and systematical errors have been combined). 
Although the dependence on the reference scale is clear at finite lattice spacing, the continuum extrapolated results are almost insensitive on this choice within our errors and they are therefore in agreement with the ones obtained in the main text.

\section{Topology}

Besides an efficient way of setting the scale, fields smoothed at non-zero flow time allows for a convenient definition of the topological charge, in terms of the straightforward discretisation of the topological charge density.

\begin{figure}[h!]
  \centering
\begin{minipage}{.48\textwidth}
  \centering
 \includegraphics[width=\linewidth]{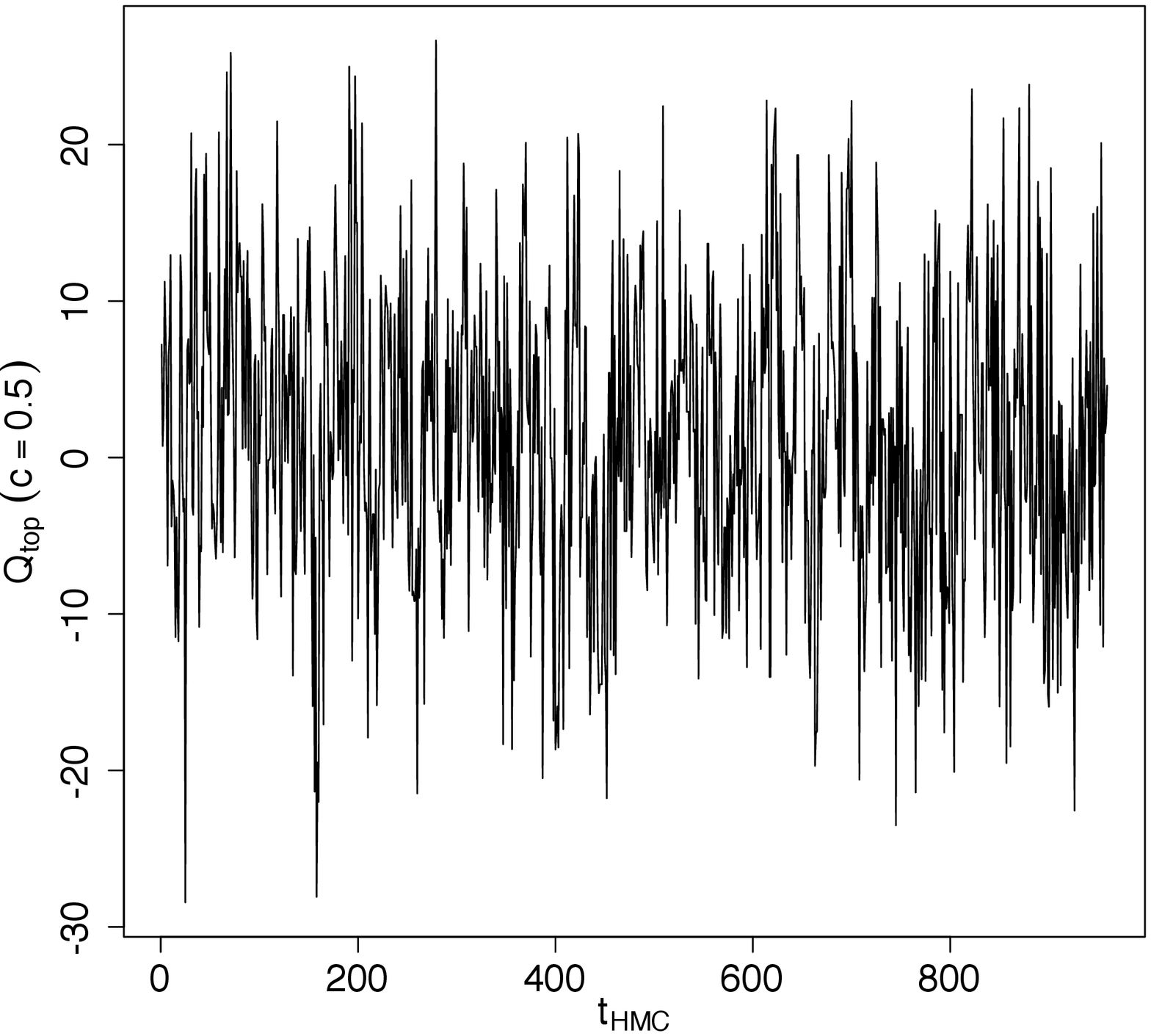}
\end{minipage}%
\hspace*{0.5cm}\begin{minipage}{.48\textwidth}
\centering
\vspace{0.35cm}
  \includegraphics[width=\linewidth]{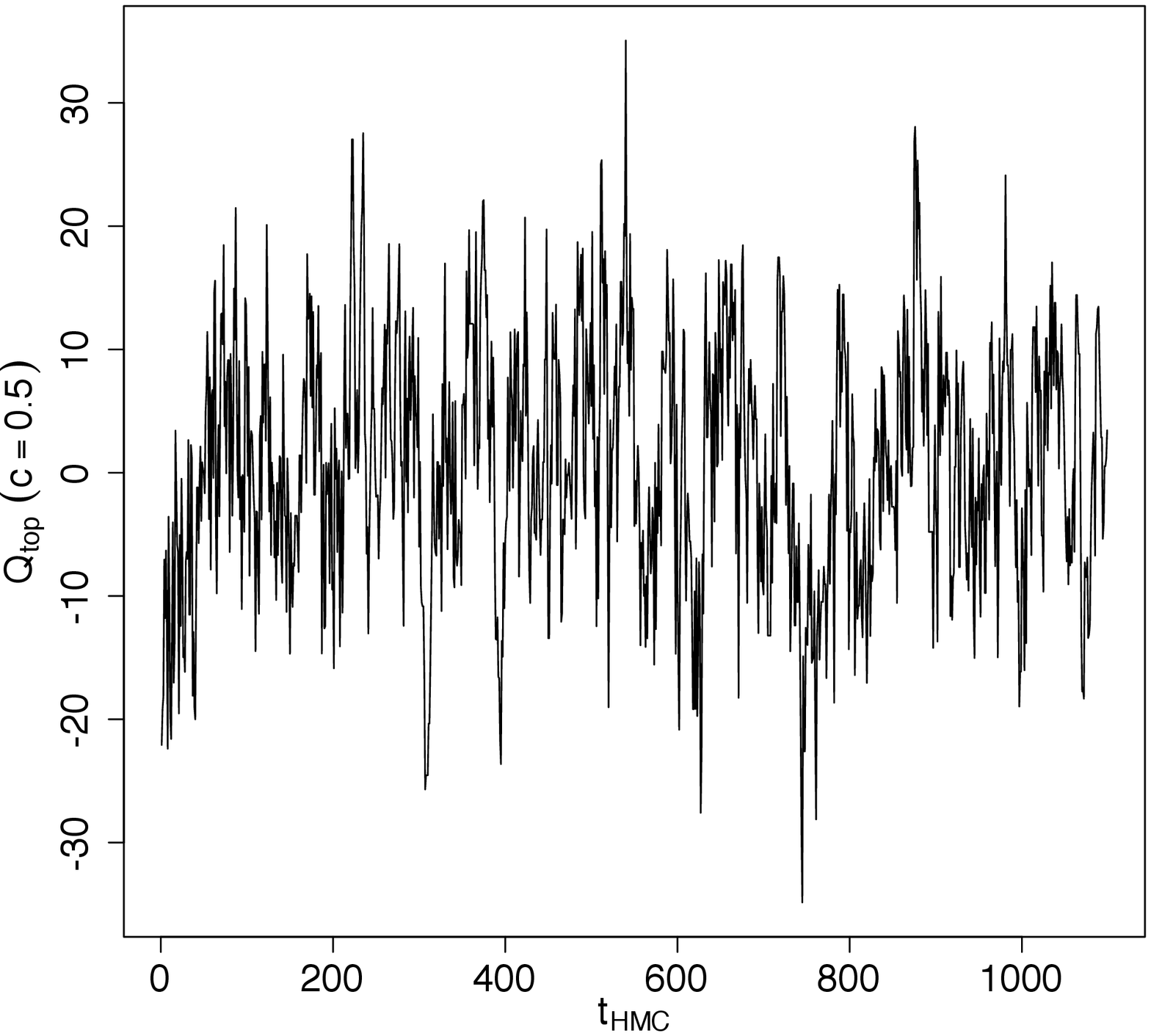}
\end{minipage}
\caption{\label{fig:Qtop_history} History of the topological charge for the most chiral run at $\beta=2.0$ (left) and $\beta=2.2$ (right) as the function of the Monte Carlo time $t_{\rm{HMC}}$.}
\end{figure}

\begin{figure}[h!]
  \centering
\begin{minipage}{.48\textwidth}
  \centering
 \includegraphics[width=\linewidth]{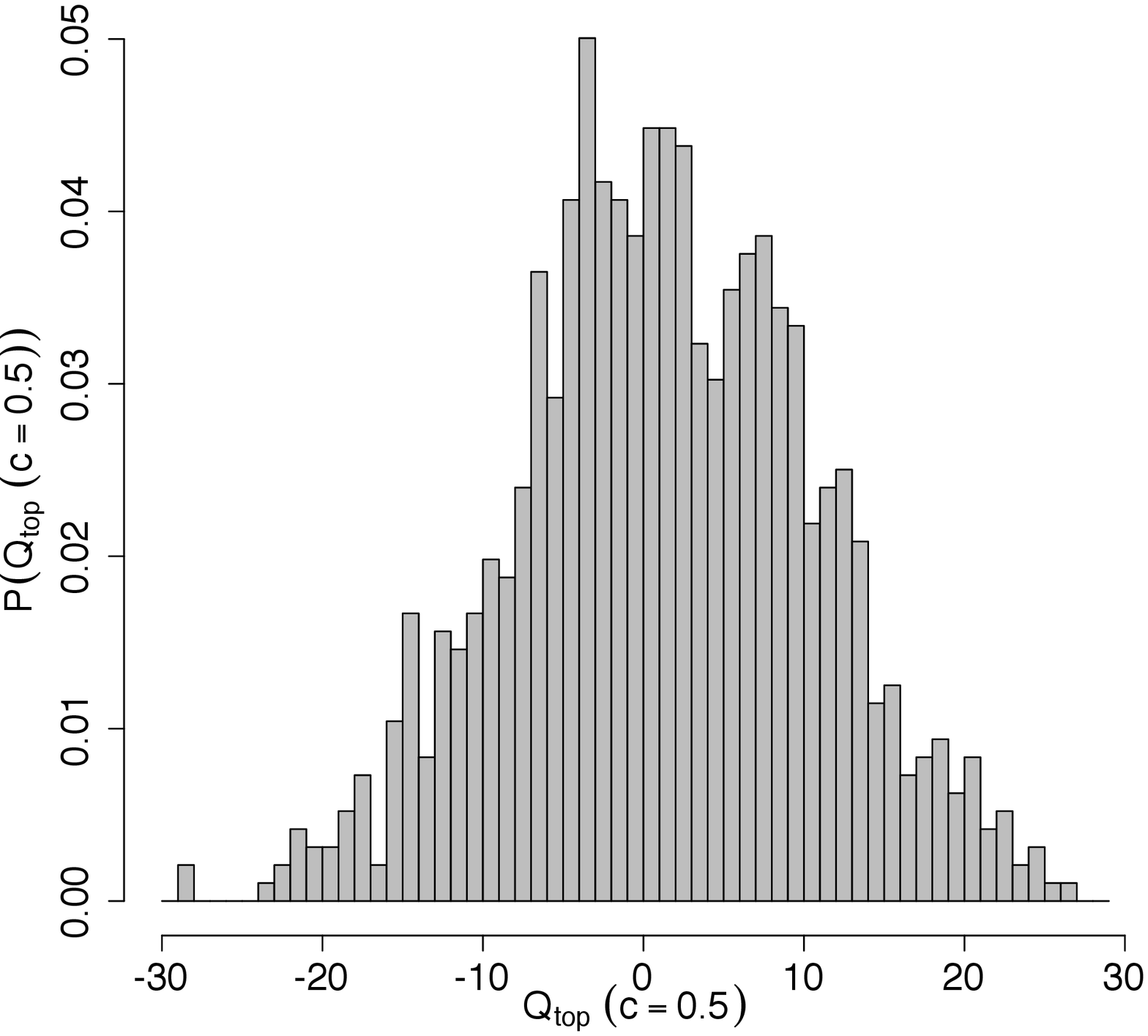}
\end{minipage}%
\hspace*{0.5cm}\begin{minipage}{.48\textwidth}
\centering
\vspace{0.35cm}
  \includegraphics[width=\linewidth]{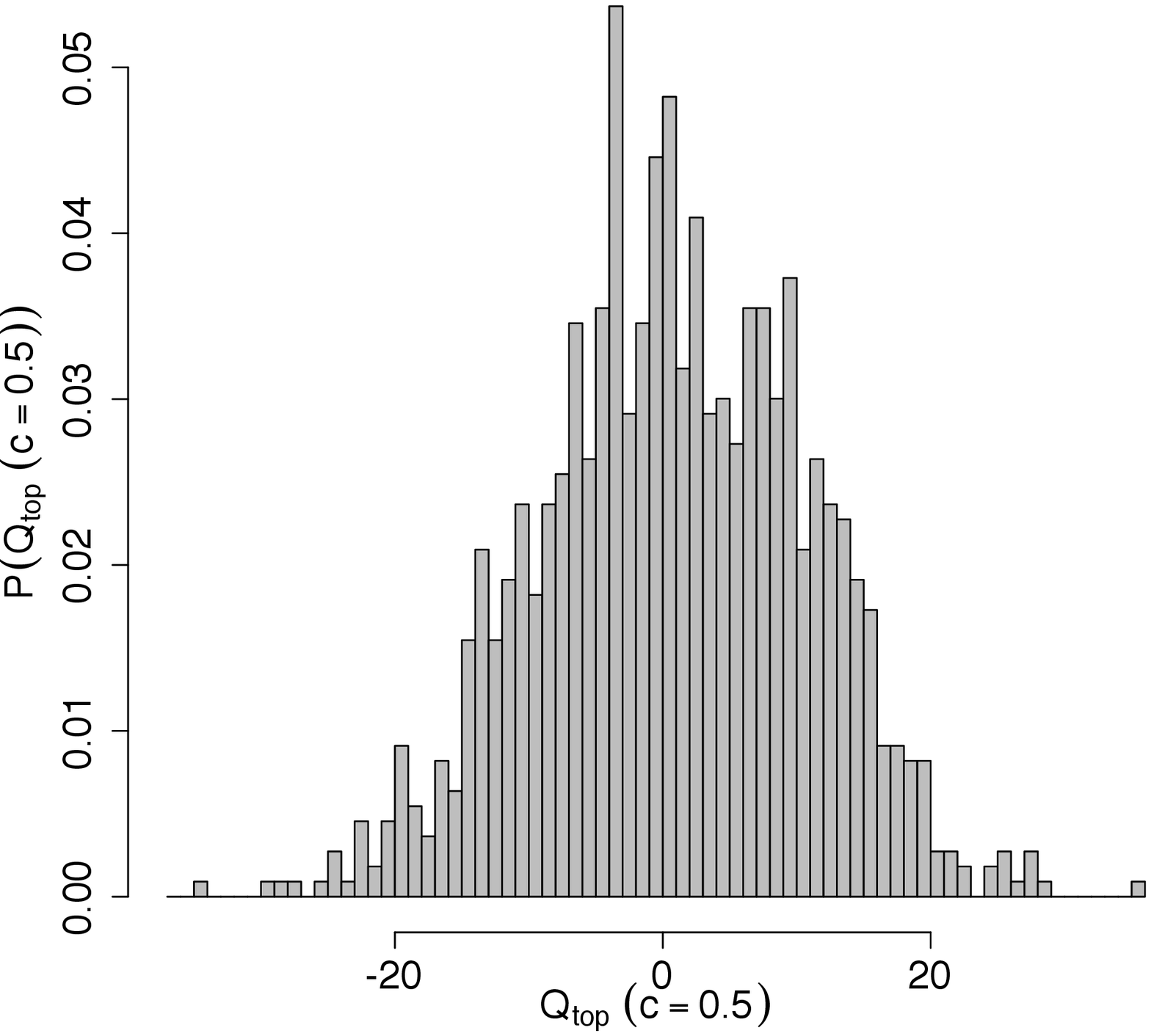}
\end{minipage}
\caption{\label{fig:Qtop_histogram} Histogram of the topological charge for the same run as in \fig{fig:Qtop_history} for  $\beta=2.0$ (left) and $\beta=2.2$ (right). }
\end{figure}

We plot in \fig{fig:Qtop_history} the topological charge as a function of the Monte Carlo time for two  $\beta$ values at the lightest quark mass for a fixed value of $c = \sqrt{8t}/L \approx 0.5$. 
In general, we observe that the average topological charge is compatible with zero for all our runs and that the fluctuations decrease with the fermion mass, as expected. 
Even if we observe larger correlation times for the topological charge at smaller quark masses, our simulations still explore all topological sectors with a good efficiency. 
The corresponding distribution of the topological charge are approximately Gaussian as shown in~\fig{fig:Qtop_histogram}.

\end{appendix}

\bibliography{paper_su2}

\end{document}